\documentclass[fleqn,usenatbib]{mnras}
\usepackage{mathptmx}

\usepackage[T1]{fontenc}
\usepackage{ae,aecompl}

\usepackage{graphicx}	
\usepackage{amsmath}	
\usepackage{amssymb}	

\newcommand{\kev}{keV}
\newcommand{\chandra}{\textit{Chandra}}

\newcommand{\bat}{\textit{Swift}-BAT}

\newcommand{\fe}{Fe~K$\alpha$}
\newcommand{\etal}{et al.}
\newcommand{\xlf}{$\phi(L)$}
\newcommand{\clf}{$\Psi(L|M_{\mathrm{h}})$}
\newcommand{\hmf}{$n(M_{\mathrm{h}})$}
\newcommand{\mh}{$M_{\mathrm{h}}$}
\newcommand{\cf}{$\xi_{\mathrm{AA}}(r)$}
\newcommand{\cfoneh}{$\xi_{\mathrm{AA}}^{1h}$}
\newcommand{\cftwoh}{$\xi_{\mathrm{AA}}^{2h}$}

\title[A New Era of Black Hole Demographics -- I. The CLF of AGNs]{Clustering,
  Cosmology and a New Era of Black Hole Demographics -- I. The Conditional Luminosity
  Function of Active Galactic Nuclei}

\author[D. R. Ballantyne]{
D. R. Ballantyne\thanks{E-mail: david.ballantyne@physics.gatech.edu}
\\
Center for Relativistic Astrophysics, School of Physics, Georgia
  Institute of Technology, 837 State Street, Atlanta, GA 30332-0430, USA\\
}

\date{Accepted XXX. Received YYY; in original form ZZZ}

\pubyear{2016}

\begin{document}
\label{firstpage}
\pagerange{\pageref{firstpage}--\pageref{lastpage}}
\maketitle

\begin{abstract}
Deep X-ray surveys have provided a comprehensive and largely unbiased
view of active galactic nuclei (AGN) evolution stretching back to $z
\sim 5$. However, it has been challenging to use the survey results to
connect this evolution to the cosmological environment that AGNs
inhabit. Exploring this connection will be crucial to understanding the
triggering mechanisms of AGNs and how these processes manifest in
observations at all wavelengths. In anticipation of upcoming
wide-field X-ray surveys that will allow quantitative analysis of AGN
environments, this paper presents a method to observationally
constrain the Conditional Luminosity Function (CLF) of AGNs at a
specific $z$. Once measured, the CLF allows the calculation of the AGN
bias, mean dark matter halo mass, AGN lifetime, halo occupation
number, and AGN correlation function -- all as a function of
luminosity. The CLF can be constrained using a measurement of the
X-ray luminosity function and the correlation length at
different luminosities. The method is illustrated at $z \approx 0$
and $0.9$ using the limited data that is currently available, and a clear luminosity dependence in the AGN bias and mean
halo mass is predicted at both $z$, supporting the idea that
there are at least two different modes of AGN triggering. In addition,
the CLF predicts that $z\approx 0.9$ quasars may be commonly hosted by haloes with
$M_{\mathrm{h}} \sim 10^{14}$~M$_{\odot}$. These `young cluster'
environments may provide the necessary interactions between gas-rich
galaxies to fuel luminous accretion. The results derived from this method will be
useful to populate AGNs of different luminosities in cosmological
simulations.  
\end{abstract}




\section{Introduction}
\label{sect:intro}
The supermassive black holes that lurk at the center of almost all
massive galaxies play an important, but still largely mysterious,
role in the formation and evolution of their host
galaxies. Significant amounts of energy and momentum can be deposited
in the central regions of a galaxy when the black hole is
accreting rapidly from its surroundings and shining as an active
galactic nucleus (AGN; \citealt{hopk05,sdh05}). The AGN phase of a galaxy may
therefore exert a significant influence on the overall size and
evolution of the galaxy \citep[e.g.,][]{dsh05,crot06,menci06,hopk08}. Unraveling the physics behind the
triggering of AGN activity and its influence on the host galaxy is an
important ingredient for a complete picture of galaxy formation and
evolution.

As the AGN phase is crucial to galaxy evolution, there has been
tremendous interest in determining a comprehensive census of AGN
activity in the Universe, and how it changes over cosmic time
\citep[e.g.,][]{bh05,tu12,mh13,ba15}. This demography of accreting black hole has its roots in the
optical quasar surveys of the 1980s \citep[e.g.,][]{hs90}, but it is now understood
that deep X-ray surveys provide the least biased determination of AGN
activity in the Universe \citep[e.g.,][]{ba15}. As a result, the last decade has
witnessed an army of X-ray
telescopes perform numerous extragalactic surveys that, when combined, are
painting a clear and consistent picture of AGN\footnote{At
  least for Compton-thin AGNs. Accreting black holes that are absorbed
  by Compton-thick material are difficult to detect outside of the local
  Universe \citep{bh05}, so their contribution and evolution remains
  largely unknown.} activity up to $z \sim 5$ \citep[e.g.,][]{ueda03,laf05,ueda14,aird10,aird15,aird15b,buch15,geo15}. Analysis of
the resulting X-ray luminosity functions (XLFs) show that AGNs underwent
strong luminosity and density evolution with time, so that the
luminosity corresponding to the peak density drops at lower
$z$ \citep[e.g.,][]{ueda14,aird15}. This evolution is likely associated with changes in the
AGN fueling and triggering physics as the Universe expanded \citep[e.g.,][]{georg09,hickox09,allev11,db12,hkb14}. 

A limitation of many of the existing deep X-ray surveys is their
narrow field of view, which limits the investigation of how AGNs are
connected to their cosmological environment. This problem becomes
especially acute as one of the leading models of AGN triggering is
mergers and interactions between galaxies \citep[e.g.,][]{san88,hernq89,kh00,dsh05,hopk05,hopk06,hopk08}. Moreover, numerical
simulations of galaxy and black hole evolution in cosmological volumes
require rigorous, quantitative constraints on not just the numbers and
luminosities of AGNs at a specific redshift, but also on how they are
distributed throughout the galaxy population. This information can be
provided by measuring the clustering of AGN as a function of
luminosity and redshift, but the wide-area surveys needed for precise
measurements are generally not available in the X-ray band \citep{caf12}. The
clustering of optical quasars has been explored for several years \citep[e.g.,][]{por04,croom05,myers07,ross09,zeh11,shen13,eft15}, but
is not able to provide an unbiased sample of black hole activity \citep[e.g.,][]{mendez16}. 

This observational landscape will be changing, however, as both
\textit{eROSITA} and \textit{Athena} will be performing wide-field
X-ray surveys that will allow precise AGN clustering measurements at
these energies \citep{kold13,kold13b}. The addition of clustering data to the
traditional suite of AGN demographic information (luminosity function,
obscuration fraction, etc.) will provide a powerful tool to
understanding how AGNs of different luminosities connect to their
large scale environments. This paper and its companion (hereafter,
Paper II; \citealt{ball16}) presents a theoretical framework, and some initial results,
of how one might exploit the potential in the new era of AGN
demographics. The underlying concept presented here is the \emph{conditional luminosity
  function} (CLF), first developed for galaxy clustering studies by
\citet{yang03} and \citet{vym03}. The CLF is a statistical connection between the AGN luminosity
function and cosmological structure that underpins the clustering
data. Constraining the CLF allows one to perform calculations in both
the traditional AGN demographics 'space' (i.e., number counts, X-ray
background spectra) and the clustering/dark-matter 'space' (i.e.,
bias, mean halo mass). This framework should also provide a method for
comparing observational data from the new wide-field X-ray surveys to
predictions of cosmological galaxy evolution simulations.

This paper provides the basic description of the
CLF framework for the application of X-ray surveys of AGNs
(Section~\ref{sect:CLF}), and then applies this analysis to AGNs at $z
\approx 0$ and $z \approx 0.9$ (Section~\ref{sect:apply}) using the
limited existing X-ray survey data with results presented in
Section~\ref{sect:results}. The implications of these results for
models and AGN triggering and evolution are discussed in
Section~\ref{sect:discuss}, while general conclusions are summarized
in the final Section. The Appendices collect several details of the
calculation procedures, as well as other supplemental results that may
be of more specialized use. Paper II shows how to apply this CLF framework
to subsets of AGNs, in particular obscured and unobscured AGNs at $z
\approx 0$ and $0.9$. A WMAP9 $\Lambda$CDM cosmology is assumed for both papers:
$h=0.7$, $\Omega_m=0.279$, $\Omega_{\Lambda}=0.721$ and
$\sigma_8=0.821$ \citep{hinshaw13}.

\section{The Conditional Luminosity Function of AGNs}
\label{sect:CLF}
This section presents an overview of the CLF and how it provides a
powerful tool to connect the evolution of AGNs to their cosmological
surroundings. Much of the mathematical description of the CLF shown
below was presented
earlier in the context of galaxy clustering \citep{yang03,vym03,van07}, but is repeated
here for completeness. Finally, the CLF framework falls under the more
general description of a Halo Occupation Distribution (HOD) model of
clustering. HOD modeling of AGN clustering has begun to be more common
in the last several years \citep{miy11,rich12,rich13,shen13}, but much of the work necessarily focuses on
luminosity-integrated quantities. As is seen below, the
luminosity-dependent CLF will be particularly important for AGN
clustering measurements expected in the near future.

\subsection{Definitions and Useful Formulas}
\label{sub:CLFintro}
We define the CLF of a specific sample of AGNs, \clf, so that
\begin{equation}
\label{eq:clf}
\phi(L)=\int_{0}^{\infty} \Psi(L|M_{\mathrm{h}}) n(M_{\mathrm{h}})
dM_{\mathrm{h}},
\end{equation}
where \xlf$=d\Phi(L)/d(\log L)$ is the XLF
in the $2$--$10$~\kev\ energy band and \hmf\ is the dark matter halo
mass function. All of these quantities are evaluated at a specific
redshift $z$, but the $z$-dependence is omitted for clarity. It is important to emphasize that \clf\ is really a
statistical description of how AGNs are distributed in haloes, where $\Psi(L|M_{\mathrm{h}})dL$ is the mean
number of AGNs in a halo of mass \mh\ with a $2$--$10$~\kev\ luminosity $L$ in the interval
$L\pm dL/2$. As such, it is not a physical quantity in the sense that
its functional form can be calculated from first principles. However,
given the appropriate datasets, the form of \clf\ can be constrained,
and then used to compute statistics of interesting physical properties of the
specified AGN population. For example, the mean number of AGNs in a
luminosity interval $[L_1,L_2]$ as a function of \mh\ is
\begin{equation}
\label{eq:avgN}
\left < N(M_{\mathrm{h}}) \right > = \int_{L_1}^{L_2}
\Psi(L|M_{\mathrm{h}}) dL,
\end{equation}
and then the average \mh\ hosting an AGN with a luminosity in this
interval is \citep{yang03}
\begin{equation}
\label{eq:avgM}
\left <M_{\mathrm{h}}(L) \right > = { \int_{0}^{\infty} M_{\mathrm{h}}
  \left < N(M_{\mathrm{h}}) \right >  n(M_{\mathrm{h}})
  dM_{\mathrm{h}} \over \int_{0}^{\infty} \left < N(M_{\mathrm{h}}) \right >  n(M_{\mathrm{h}})
  dM_{\mathrm{h}}}
\end{equation}
\begin{equation*}
\phantom{\left <M_{\mathrm{h}}(L) \right >} = {1 \over \phi(L)} \int_{0}^{\infty} M_{\mathrm{h}}
  \left < N(M_{\mathrm{h}}) \right >  n(M_{\mathrm{h}})
  dM_{\mathrm{h}}.
\end{equation*}
Tracking these quantities as a function of $z$ will provide important
insights into how AGNs evolve and their dependence on their
environment.

The CLF also allows a direct estimate of the AGN lifetime or
duty cycle as a function of luminosity. \citet{mw01} showed that an estimate
of the AGN lifetime would be
\begin{equation}
\label{eq:duty}
\tau_{\mathrm{AGN}}(L) = {\nu_{\mathrm{AGN}}(L,M_{\mathrm{h}}) \over \nu(M_{\mathrm{h}})} \tau_{\mathrm{Hubble}},
\end{equation}
where $\nu_{\mathrm{AGN}}(L,M_{\mathrm{h}})$ is the space density of
AGNs with luminosity $L$ present in haloes with mass \mh,
$\nu(M_{\mathrm{h}})=n(M_{\mathrm{h}})dM_{\mathrm{h}}$ is the space density of those haloes, and
$\tau_{\mathrm{Hubble}}$ is the Hubble time at the $z$ of
interest. Using the CLF of AGNs, the numerator can be replaced with
$\Psi(L|M_{\mathrm{h}}) n_{\mathrm{h}} dL dM_{\mathrm{h}}$, and, after
replacing the halo mass with $\left <M_{\mathrm{h}}(L) \right >$,
$\tau_{\mathrm{AGN}}(L)$ can be simply written as
\begin{equation}
\label{eq:time}
\tau_{\mathrm{AGN}}(L) = (\Psi(L|\left <M_{\mathrm{h}}(L) \right >) dL) \tau_{\mathrm{Hubble}}.
\end{equation}

\subsection{How To Determine the AGN CLF}
\label{sub:CLFhowto}
As the CLF is purely a statistical description of the AGN population,
it has to be constrained from measurements. The first important
dataset is the XLF (i.e., the left-hand side of Eq.~\ref{eq:clf}), but
this is not enough on its own, as even after specifying the halo mass
function \hmf, there is an infinite number of possible \clf\ that can
result in the measured XLF. Thus, additional information that depends
on both luminosity and halo mass is required to determine the shape of
the CLF. As shown by \citet{yang03} and \citet{vym03} for galaxies, the variation of correlation
lengths with luminosity, $r_0(L)$, can provide the needed
measurements. 

The correlation lengths come from measurements of the AGN two-point
correlation function \cf\ which measures the probability above random chance that
two random AGNs within a volume element $dV$ will be separated by a physical distance $r$:
\begin{equation}
\label{eq:xi}
dP=\nu[1+\xi_{\mathrm{AA}}(r)]dV,
\end{equation}
where $\nu$ is the mean number density of AGNs in $dV$ \citep{peebles80}.
Observations of \cf\ at both the optical and X-ray wavelengths show
that \cf\ is well
approximated by a power-law, $\xi_{\mathrm{AA}} = (r/r_0)^{-\gamma}$,
where $r_0$, the correlation length, is defined so that
$\xi_{\mathrm{AA}}(r_0)=1$ \citep[e.g.,][]{caf12}. The \cf\ will depend on both $z$,
luminosity and, possibly, AGN type, but for clarity only the spatial
dependence is indicated. Observations typically measure the projected
correlation function $w(r_p)$ (where $r_p$ is the projected distance
between a pair of AGNs in the plane of the sky), which is related to
the 'real-space' \cf\ via \citep{dp83}
\begin{equation}
\label{eq:projcf}
w(r_p) = 2 \int_{r_p}^{\infty} {r \xi_{\mathrm{AA}}(r) \over \sqrt{r^2
    - r_p^2}} dr.
\end{equation}
Assuming that the true correlation function is $\xi_{\mathrm{AA}} =
(r/r_0)^{-\gamma}$ then the observed $w(r_p)$ gives an estimate of
$r_0$ \citep{dp83}.

In HOD modeling it is useful to decompose the total
\cf\ into a `1-halo' term, \cfoneh, and a `2-halo'
term, \cftwoh\ \citep[e.g.,][]{yang03}:
\begin{equation}
\label{eq:1h2h}
\xi_{\mathrm{AA}}(r)=\xi_{\mathrm{AA}}^{1h}(r)+\xi_{\mathrm{AA}}^{2h}(r).
\end{equation}
The `1-halo' term represents the contribution to the \cf\ from pairs
of AGNs in a single halo, sometimes called satellite AGNs. Determining this term theoretically requires
specifying a model of how galaxies and AGNs are distributed within a
halo. The transition between the two terms occurs at $r \sim
1$~h$^{-1}$Mpc $ < r_0$ for most AGN correlation functions \citep{caf12}, so the
`1-halo' term will be ignored for determining the CLF  (similarly, satellite AGNs were not
considered in models developed by \citet{croton09} and
\citealt{cw13}). Thus,
$\xi_{\mathrm{AA}} \approx \xi_{\mathrm{AA}}^{2h}$. 

As the `2-halo' term denotes the clustering of AGNs in separate haloes,
then they can clearly be related to the clustering of the haloes
themselves,
\begin{equation}
\label{eq:xi2h}
\xi_{\mathrm{AA}}^{2h}(r) \approx \bar{b}_A^2 \xi_{\mathrm{dm}}^{2h},
\end{equation}
where $\xi_{\mathrm{dm}}^{2h}$ is the `two-halo' contribution to the total
dark matter correlation function (defined as $\xi_{\mathrm{dm}}^{2h}=\xi_{\mathrm{dm}}-\xi_{\mathrm{dm}}^{1h}$), and 
\begin{equation}
\label{eq:bbar}
\bar{b}_A={\int_{0}^{\infty} n(M_{\mathrm{h}}) \left <
    N(M_{\mathrm{h}}) \right > b(M_{\mathrm{h}}) dM_{\mathrm{h}} \over
  \int_{0}^{\infty} n(M_{\mathrm{h}}) \left <
    N(M_{\mathrm{h}}) \right > dM_{\mathrm{h}}}
\end{equation}
is the mean AGN bias at a specified $z$. Since $\left <
N(M_{\mathrm{h}}) \right >$ can be computed for any interval in
luminosity (Eq.~\ref{eq:avgN}), the AGN bias (and therefore the
correlation lengths) can also be calculated as a function of
luminosity. The $b(M_{\mathrm{h}})$ factor in Eq.~\ref{eq:bbar} is the bias of dark
matter haloes to the underlying dark matter distribution and depends on the
definition of `halo' that is adopted.

At this point the path toward determining the CLF for a sample of AGNs
is laid out once a specific cosmological model is chosen. Given a parameterization of the CLF, and the XLF and $r_0(L)$
data for the sample in question, predictions for the AGN correlation
lengths and XLF can be computed using the above equations and compared
with the data. A fitting technique can then be used to constrain the
parameters of the CLF. As seen above, the procedure involves several
quantities that depend only on the cosmological model. The details of
how these are computed in this paper can be found in Appendix~\ref{app:cosmo}. The
next Section applies this technique to two samples of AGNs, one at $z
\approx 0$ and one at $z \approx 0.9$.

\section{Application to AGN Data}
\label{sect:apply}
To determine the CLF of a sample of AGNs using the above framework
requires measurements of both the XLF and $r_0(L)$ of the objects in
the sample. Due to the relatively small survey areas probed in the
$2$--$10$~\kev\ band, this last quantity has been rarely
measured. However, there are determinations of $r_0(L)$ at two
different redshifts ($z \approx 0$ and $z \approx 0.9$) which, despite
their relatively narrow luminosity range, can be
used to illustrate the insights into AGN evolution gained from the CLF
methodology. 

In the calculations described below, rest-frame $2$--$10$~\kev\ luminosities,
$L$, are computed on a grid from $\log (L/\mathrm{erg\ s^{-1}})=41.11$
to $48$ in 531 equally spaced logarithmic steps. The grid of halo
masses spans $\log (M_{\mathrm{h}}/\mathrm{M_{\odot}})=3$ to $16.473$
in 1000 steps. For computational simplicity, Eq.~\ref{eq:halo} is used
to generate a grid of radii from the halo masses on which to calculate correlation
functions. This grid extends from $\sim 1$~pc to $\sim 57$~Mpc at $z=0$,
but, due to the increase in density at higher $z$, the maximum radius
at $z=0.9$ is $\approx 20$~Mpc. These upper limits in radii cause the
predicted $w(r_p)$ to roll off at $r_p \approx 10$~Mpc (see
Eq.~\ref{eq:projcf}); however, most of the $w(r_p)$ data we compare to
are at much smaller projected radii and the predictions of $r_0(L)$
are independent of the $w(r_p)$ calculation. Therefore, given the
appropriate datasets, this setup allows for a well-determined
measurement of the AGN CLF.

\subsection{The $z \approx 0$ Dataset}
\label{sub:z0data}
For AGNs in the local Universe there is a measurement of $r_0$ in two
different luminosity ranges from the 3-year \bat\
catalog \citep{cap10}. The high luminosity value ($r_0 \approx
20$~Mpc) is likely overestimated, as it suggests that local quasars
are found in cluster-sized haloes, in contrast to observations
\citep{kauff04,pb06,mart13}. Nevertheless, it is still useful to consider these data
because the lower-luminosity point is more accurate, and any
luminosity-dependence derived from the CLF model may be instructive. 

As a further constraint on the CLF, we use the 6
$w(r_p)$ points at $2~h^{-1}\mathrm{Mpc} \la r_p \la 14$~$h^{-1}$Mpc
calculated by \citet{cap10} which use the entire luminosity range of their
sample. Values of $w(r_p)$ at larger radii
are omitted from the fits because of the roll-over
described above (the data at these projected radii also lie
significantly above the best fit power-law to $w(r_p)$ and thus may
be overestimated). The data at smaller projected radii will have
a non-negligible contribution from $\xi_{\mathrm{AA}}^{1h}(r)$ which is
not modeled by the CLF. 

The CLF is determined for $2$--$10$~keV luminosities, while the \citet{cap10}
data is measured in the \bat\ $15$--$55$~keV band. An average AGN
spectral energy distribution (SED) consisting of a cutoff power-law and
reflection spectrum is used to convert between the two bands. SEDs
are computed for a range of photon indices ($\Gamma=1.4$--$2.4$) and
then Gaussian averaged with a mean $\Gamma=1.9$ and $\sigma=0.3$ to
obtain the final SED. An e-folding energy of $250$~\kev\ is used for
all spectra. Reflection is included using the \textsc{pexmon} model \citep{nan07}
 available in XSPEC \citep{arn96} and is added to each power-law prior to averaging
so that the \fe\ equivalent width is $120$~eV. The final SED is
consistent with observations of many samples of local AGN \citep[e.g.,][]{burlon11} as
well as the average spectrum derived by fitting the local XLFs \citep{ball14}. The
total luminosity range covered by the \citet{cap10} $w(r_p)$ data is estimated
to be $42.4 < \log (L_{\mathrm{15-55\ keV}}/(\mathrm{erg\ s^{-1}})) <
43.9$ which, according to the AGN template, corresponds to $42.25 <
\log (L_{\mathrm{2-10\ keV}}/(\mathrm{erg\ s^{-1}})) < 43.75$.

The $z \approx 0$ $2$--$10$~keV XLF data is taken from \citet{ueda14} and
\citet{dc08}. The reason for including the \citet{dc08} estimate (which is a
de-evolved measurement and therefore model dependent) is that X-ray
background synthesis models employing the \citet{ueda14} XLF overpredict the
local \bat\ number counts \citep{aird15b,harr15}. Noting this problem, \citet{ball14} provided a
corrected local $2$--$10$~keV XLF that lies between the \citet{ueda14}
and \citet{dc08} data, but will also fit the \bat\ number counts. Thus, a more
accurate measurement of the local AGN population is obtained by
including both the \citet{ueda14} and \citet{dc08} XLF datapoints. A negative
consequence of this decision is that, because the two XLFs are offset
for each other, a single XLF model will be a poor statistical fit to
these data. Combining these two XLF measurements with the correlation
measurements described above yields 32 datapoints available for fitting
the CLF.

Now that the necessary data are collected, a form for the AGN CLF must
be decided upon and parameterized. As the CLF is statistical in
nature, there is freedom to fix the form so that it `makes sense'. For
example, the XLF at most redshifts is well described by a broken
power-law, so integrating the CLF over halo mass (e.g.,
Eq.~\ref{eq:clf}) should give a broken power-law. Similarly, HOD modeling of
AGNs indicate that $\left <N(M_{\mathrm{h}}) \right >$ also has
a roughly power-law shape \citep{chatt13,lea15}, so integrating the CLF over luminosity (e.g.,
Eq.~\ref{eq:avgN}) also should yield a power-law. This power-law,
however, cannot rise indefinitely, so a cutoff halo mass that can
depend on AGN luminosity should be included. These considerations lead
to the following form for the AGN CLF:
\begin{equation}
\label{eq:clfform}
\Psi(L|M_{\mathrm{h}}) = \left( {M_{\mathrm{h}} \over M_{\ast}}
\right )^{a} e^{-M_{\mathrm{h}}/M_{\mathrm{cut}}} f(L)
\end{equation}
where
\begin{equation}
\label{eq:mcut}
M_{\mathrm{cut}}=\left( {L \over L_{\ast}} \right)^c M_{N},
\end{equation}
and
\begin{equation}
\label{eq:brokenpower}
f(L)= \left \{ \begin{array}{ll}
\left( {L \over
  L_{\ast}} \right)^{-0.96} & \mbox{if $L < L_{\ast}$} \\
\left( {L \over
  L_{\ast}} \right)^{-\beta} & \mbox{otherwise.}
\end{array} \right.
\end{equation}
Fixing the low luminosity slope at $-0.96$ (equal to the
low-luminosity slope of the \citealt{ueda14} XLF) leaves 6 free parameters to constrain by fitting the
XLF and correlation data: $a$, $M_{\ast}$, $\beta$, $L_{\ast}$, $c$, and
$M_{N}$. We can simplify the process slightly by noting that, although
the mass function \hmf\ is defined down to $1000$~M$_{\odot}$, galaxies
(and therefore AGNs) are not typically found in haloes with
$M_{\mathrm{h}} \la 10^9$~M$_{\odot}$ \citep[e.g.,][]{efst92,ben02,nick11,shen14}; thus,
$\Psi(L|M_{\mathrm{h}})=0$ for those masses. Other assumed forms for the CLF (e.g., log-normals, power-laws,
exponentials) did not provide acceptable fits to the XLF data. While
more complex forms could provide equally good fits to the
data, the parameterization shown above appears to be the simplest CLF
description supported by the available data. Of course, as additional
$r_0(L)$ data becomes available, particularly at higher luminosities,
this parameterization may have to be revised.

\subsection{The $z \approx 0.9$ Dataset}
\label{sub:z09data}
\citet{kou13} measured $r_0(L)$ in two luminosity ranges for AGNs at $z
\sim 0.9$--$1$ in five different deep X-ray surveys, yielding 10
different measurements of $r_0(L)$. The values of $r_0$ determined
from the \chandra\ Deep Field South are much larger than the other
fields (likely due to the large supercluster in the field) and is
omitted from the CLF fits. Specifically, we use the $r_0$ data in the
CLF model, and not the `corrected' $r_{0,c}$ values listed by \citet{kou13}
that attempts to account for the finite area of the surveys. In
practice, these two estimates of the correlation length are almost
always consistent within the errors. As with the $z \approx 0$ field, the $r_0(L)$ data is augmented with
the $w(r_p)$ profile measured by \citet{kou13} using their entire $z \approx
0.9$ sample with the measurements at the largest and smallest $r_p$
omitted in the calculations for the reasons described
above. Finally, \citet{kou13} work with luminosities in the
$0.5$--$8$~\kev\ band, and the same AGN SED described for the $z
\approx 0$ data is used to convert these into the
$2$--$10$~\kev\ band. Therefore, the \citet{kou13} $w(r_p)$ data, that spans $41.25
\la \log (L_{\mathrm{0.5-8\ keV}}/(\mathrm{erg\ s^{-1}})) \la 44.5$,
converts to the range $41.1 < \log
(L_{\mathrm{2-10\ keV}}/(\mathrm{erg\ s^{-1}})) < 44.3$. The XLF data for the CLF fitting is taken from the $z=0.8$--$1.0$
panel in the \citet{ueda14} measurement of the evolving XLF. Combining the \citet{ueda14} XLF data points with
the \citet{kou13} correlation function data gives a total of 26 data points
with which to constrain Eqs.~\ref{eq:clfform}--~\ref{eq:brokenpower}.

Determining the CLF at this redshift involves calculations of \hmf\ and
$\xi_{\mathrm{dm}}$ at one $z$ (namely, $z=0.9$) while the data that
is constraining the CLF has been gathered over a range of redshift
(e.g., $z \approx 0.7$--$1.1$ for the correlation data; \citealt{kou13}). In
principle, this mismatch may lead to an inaccurate measurement of
AGN properties at $z=0.9$. As shown by \citet{van07} and described in
Appendix~\ref{app:light-cone}, the CLF equations can be adapted so
that an `effective' \hmf\ and $\xi_{\mathrm{dm}}^{2h}$, defined over the
necessary $z$ range, is used in place of the ones calculated at a
single redshift. In practice, this correction is too small for the
redshift range applicable here, and so is not used for the current
calculation.  

\subsection{Fitting the Data}
\label{sub:fitting}
For both redshifts, the 6 CLF parameters are determined by minimizing
the total $\chi^2$ from all data points using the Metropolis
algorithm \citep{metro53} to ensure
that the true global minimum is found. To obtain $\chi^2/$dof$ \approx
1$ (dof=degree of freedom) we used the larger of the two errorbars in
the XLF datapoints, but otherwise used the smallest ones for the
$w(r_p)$ and $r_0$ data. This procedure assumes that all the errorbars
are independent and Gaussian distributed. The best fitting CLF parameters
are shown in Table~\ref{table:clfs}, where the uncertainties are the $90$\% confidence
level for one degree of freedom (i.e., a $\Delta \chi^2=2.71$
criterion). Due to including two non-compatible measurements of the
$z\approx 0$ XLF, the $\chi^2$ of the CLF fit is formally
poor. Fitting only to the \citet{ueda14} XLF yields a $\chi^2/$dof$=0.96$;
however, the quantities derived from the CLF will be more accurate
after including the \citet{dc08} XLF estimate.

\begin{table*}
\centering
\caption{Best-fit AGN CLF parameters at $z \approx 0$ and $z \approx 0.9$
  (Eqs.~\ref{eq:clfform}--\ref{eq:brokenpower}). The error-bars are calculated using a
  $\Delta \chi^2=2.71$ criterion (i.e., a 90\% confidence level for
  the parameter of interest).}
\label{table:clfs}
\begin{tabular}{lccccccc}
\hline
Redshift & $\chi^2/$dof & $a$ & $\log (M_{\ast}/$M$_{\odot})$ & $\beta$ &
$\log (L_{\ast}/$erg s$^{-1})$ & $c$ & $\log
(M_{\mathrm{N}}/$M$_{\odot})$\\
\hline
$0$ & $58.04/26$ & $2.20\pm 0.05$ & $12.51\pm 0.04$ &
$2.57^{+0.05}_{-0.04}$ & $41.79\pm 0.05$ & $1.27^{+0.07}_{-0.06}$ &
$12.33\pm 0.08$ \\
$0.9$ & $12.77/20$ & $2.96^{+0.08}_{-0.10}$ & $14.43^{+0.04}_{-0.03}$ &
$3.26^{+0.88}_{-0.42}$ & $44.47\pm 0.05$ &
$0.59^{+0.05}_{-0.04}$ & $15.20^{+0.07}_{-0.08}$ \\
\hline
\end{tabular}
\end{table*} 
As the CLF is a statistical quantity, one has to be careful not to overinterpret the numbers
in the Table. Nevertheless, there are some trends that one can glean
from examining the CLF fits. For example, the overall luminosity
scale increase with $z$, as observed in the evolution of the XLF \citep[e.g.,][]{ueda14,aird15}. The
mass scale, as measured by $M_{\ast}$ and $M_{\mathrm{N}}$ also increases, although this is
not typically infered from X-ray clustering measurements
\citep{caf12}. Interestingly, the parameters that quantifies the
coupling between the halo mass and the AGN luminosity ($c$) is $\approx
2\times$ lower at $z \approx 0.9$ than at $z \approx 0$.  

\begin{figure*}
\centering
\includegraphics[width=0.33\textwidth]{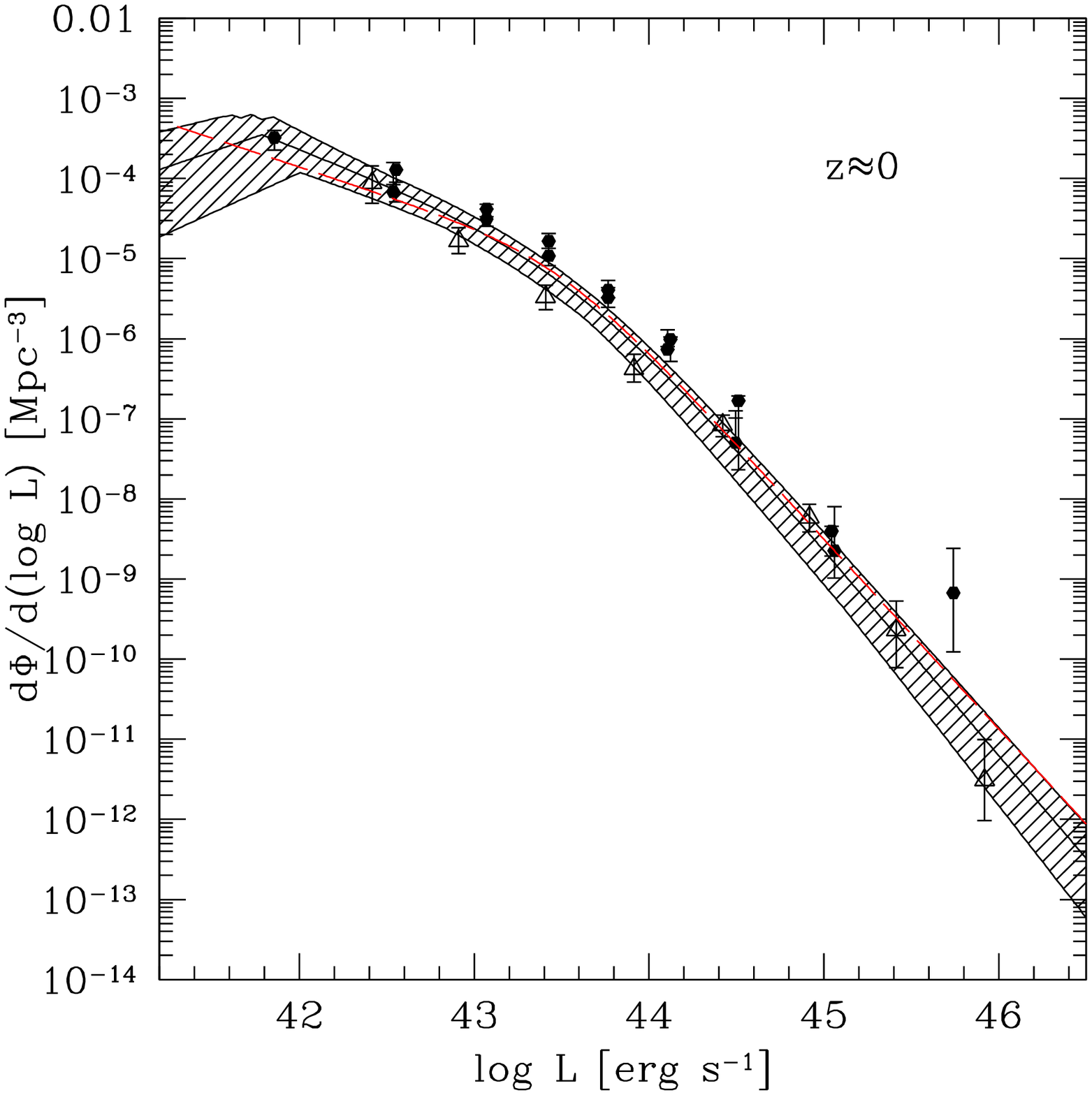}
\includegraphics[width=0.33\textwidth]{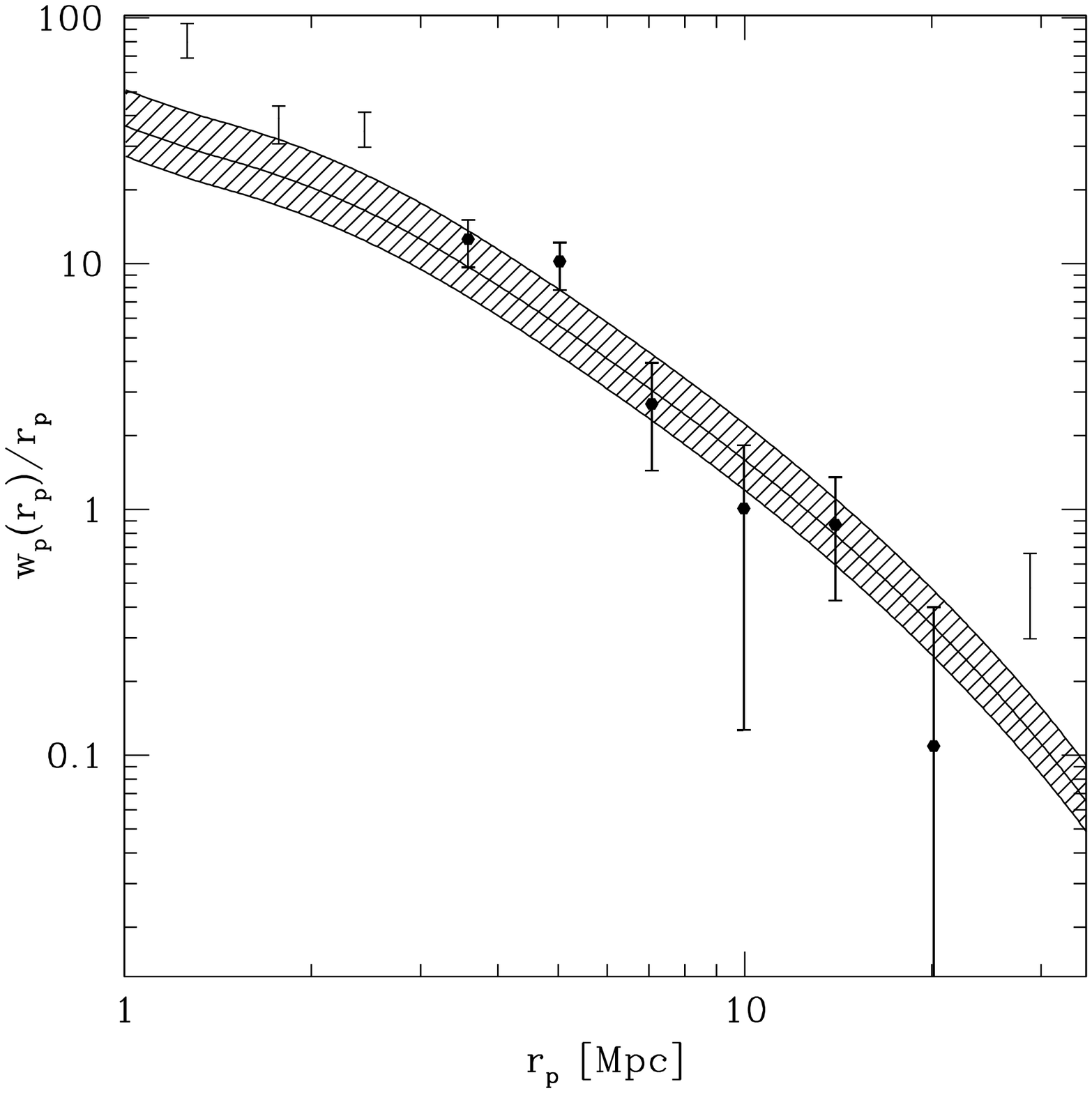}
\includegraphics[width=0.33\textwidth]{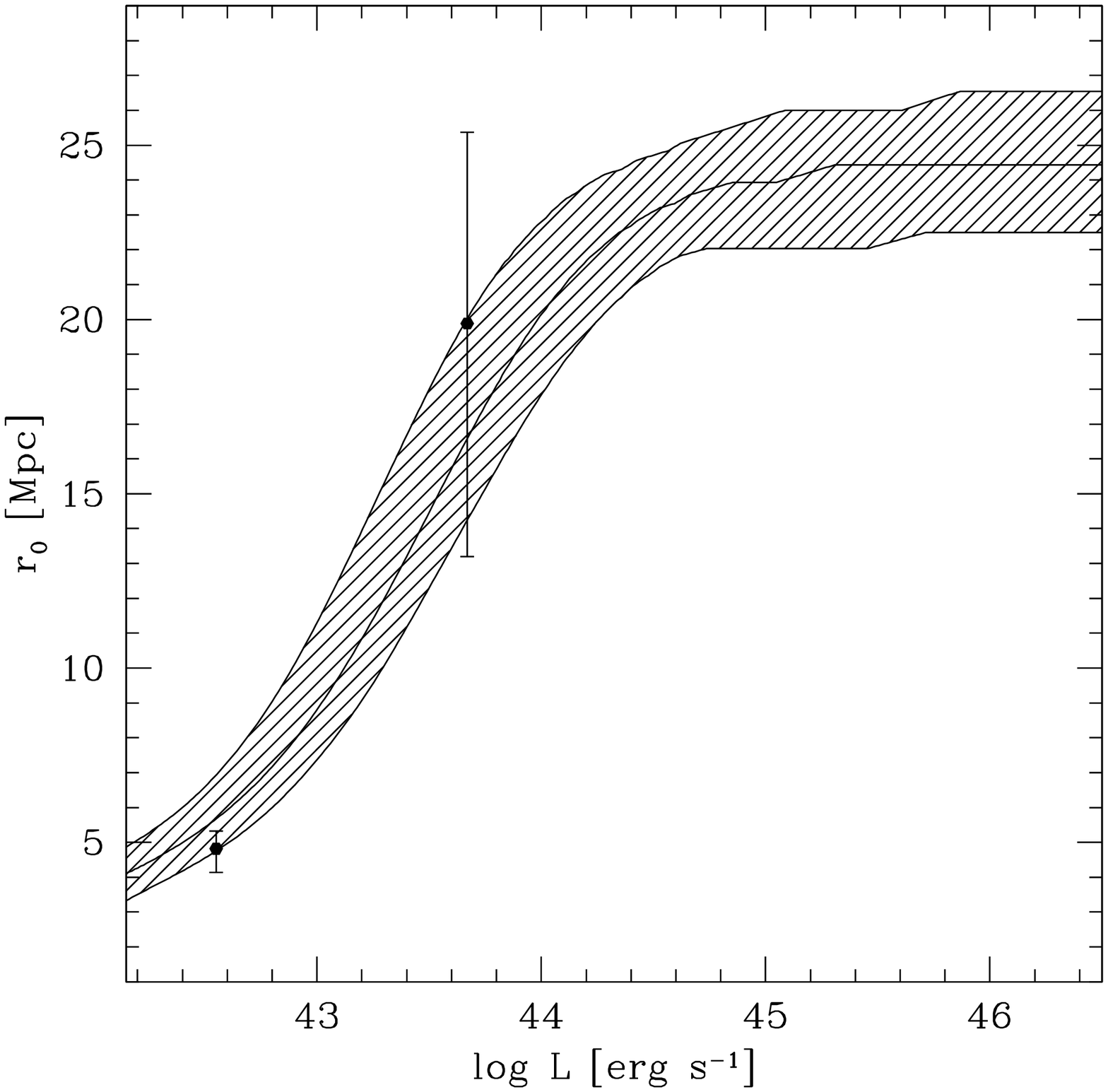}
\caption{Fits to the $z\approx 0$ XLF (left), $w(r_p)$ (middle) and
  $r_0$ (right) data that determined the AGN CLF at this redshift. In
  all plots the solid line indicates the prediction from the
  best-fit CLF (top row of Table~\ref{table:clfs}), and the hatched
  region shows the 95\% confidence region on this quantity. In the XLF
plot, the solid points denotes data taken from the paper of \citet{ueda14} and
the triangles are the \citet{dc08} estimate of the local LF. The red dashed line
in the XLF panel plots the best-fit $z=0$ XLF determined by \citet{ball14} from
fitting the local XLFs in multiple energy bands. The $w(r_p)$ and $r_0$
data of \citet{cap10} are shown in the other two panels, with the solid points
indicating ones that are actually used in the CLF fit (the other
points are omitted from the fit; see text for details). }
\label{fig:fitsz0}
\end{figure*}
\begin{figure*}
\centering
\includegraphics[width=0.33\textwidth]{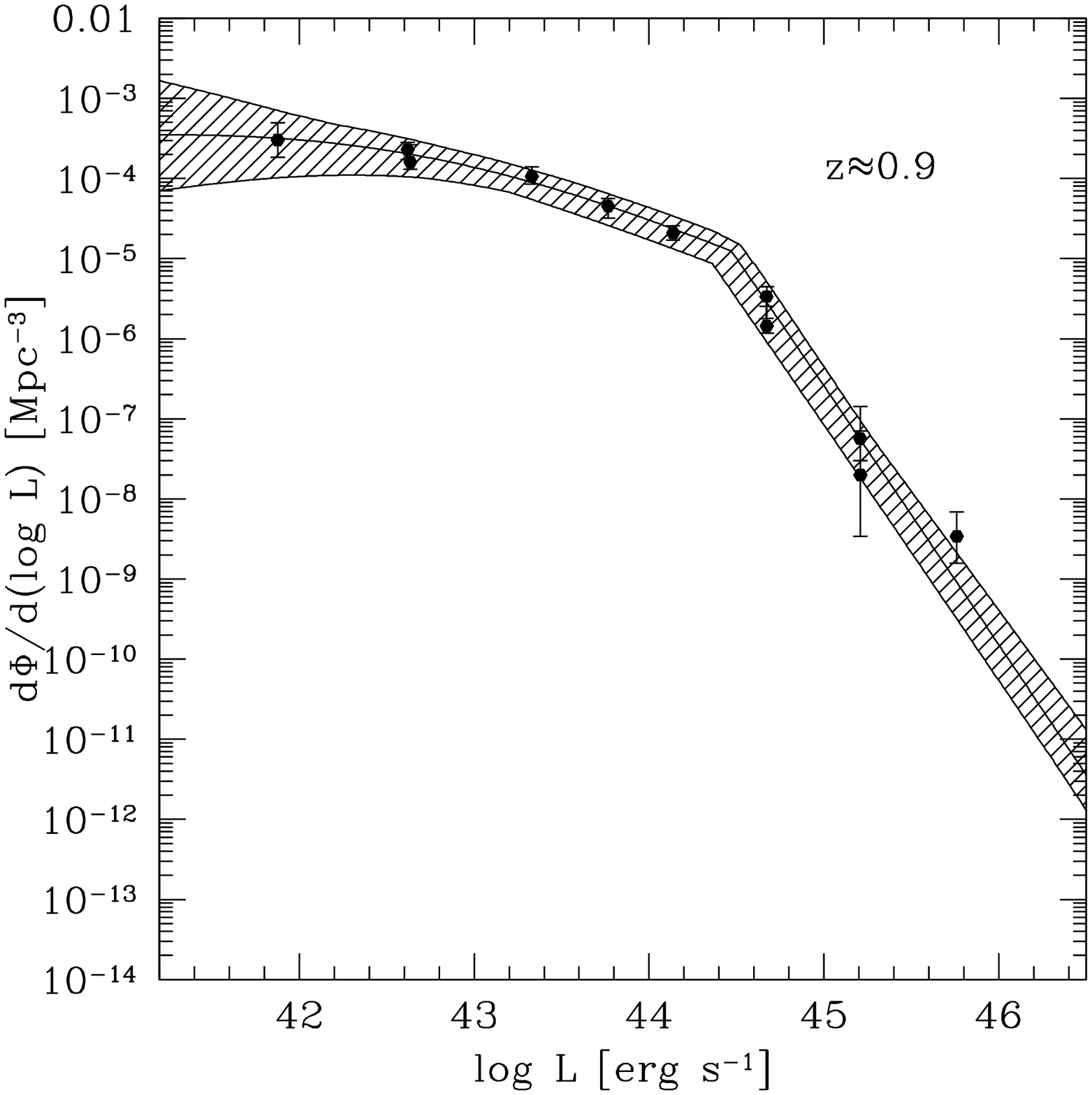}
\includegraphics[width=0.33\textwidth]{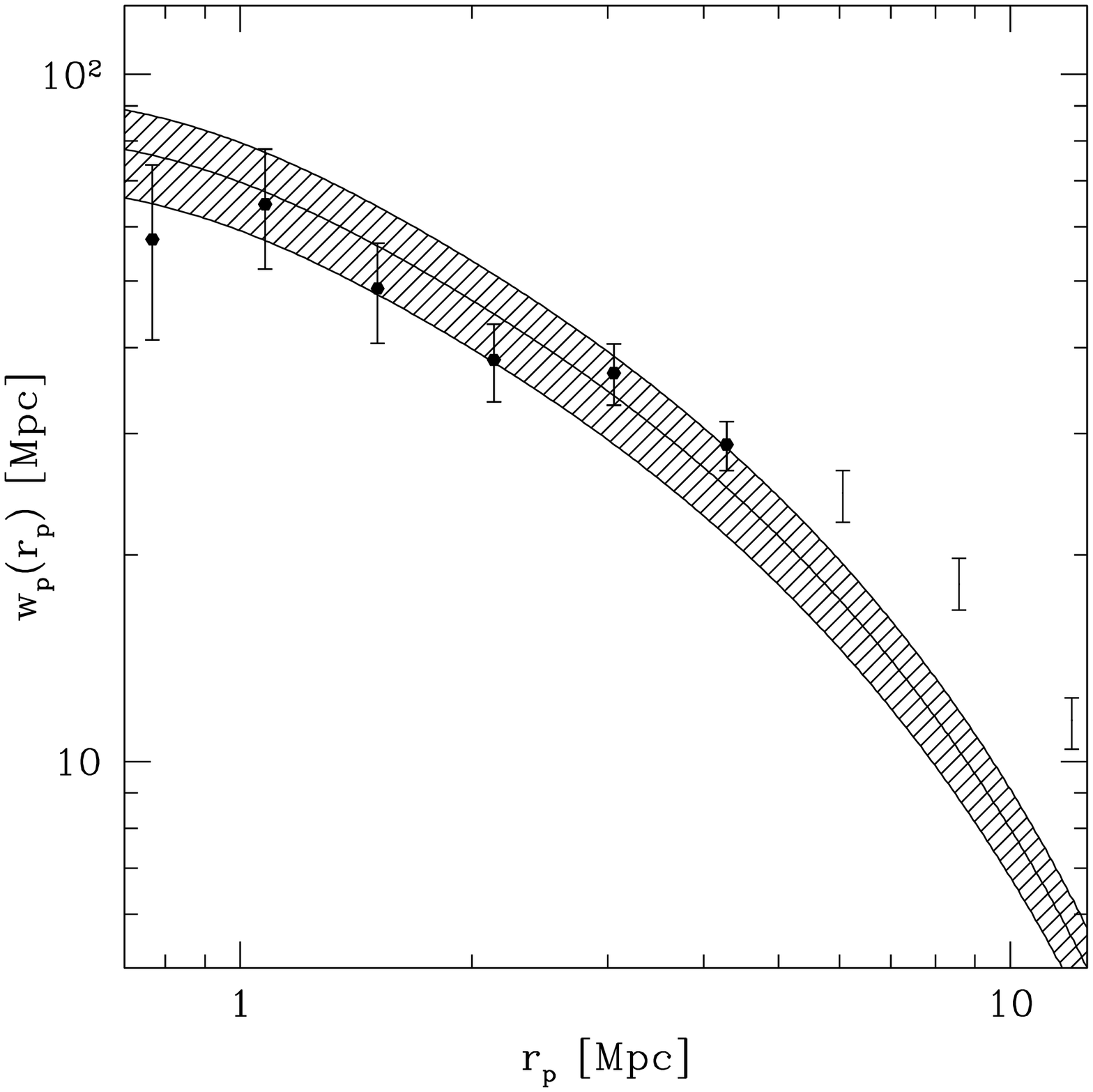}
\includegraphics[width=0.33\textwidth]{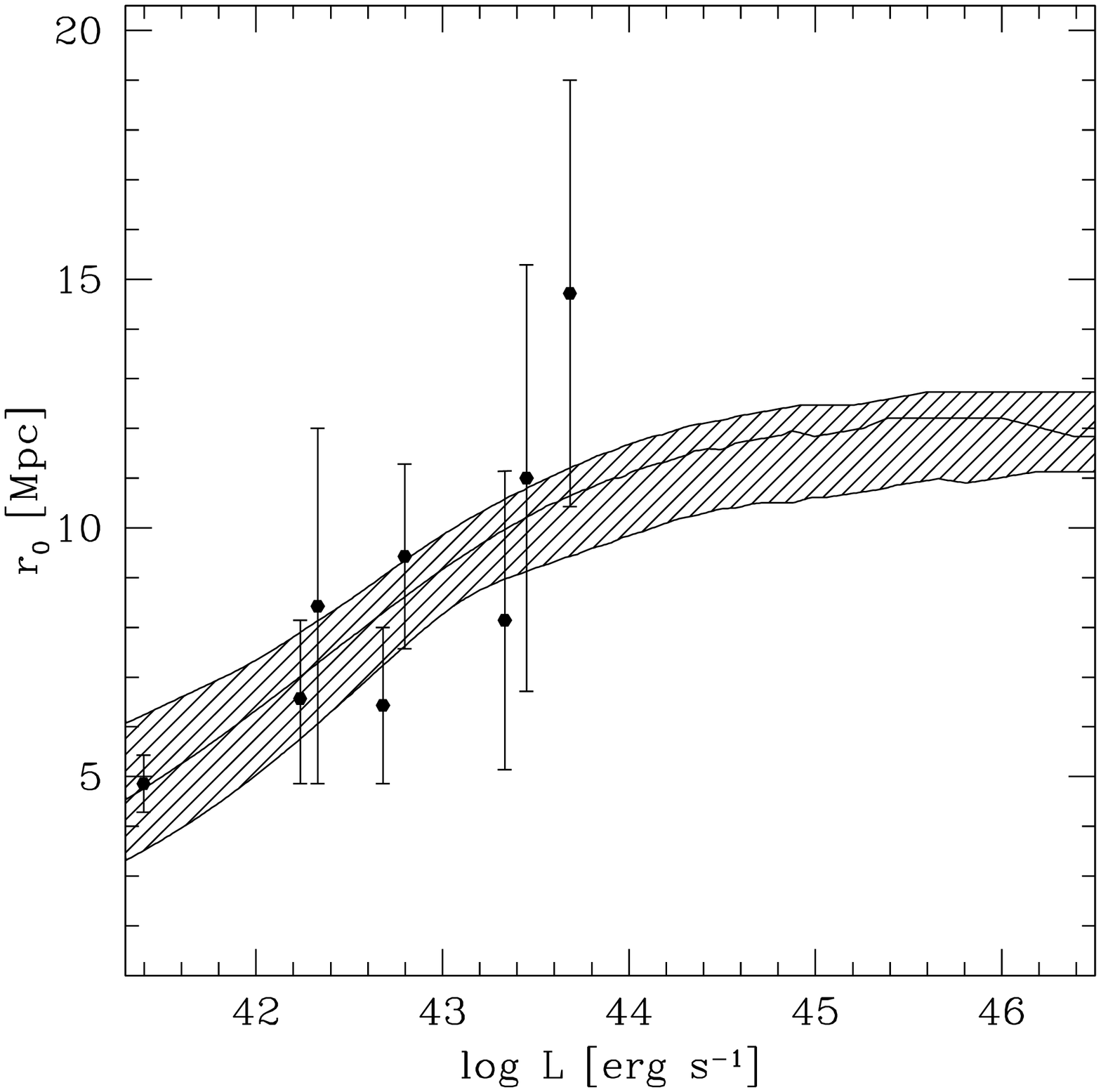}
\caption{As in Fig.~\ref{fig:fitsz0}, but at $z\approx 0.9$. In this
  case, the $w(r_p)$ and $r_0$ data are taken from \citet{kou13}. The
  parameters of the best-fit CLF are listed in the bottom row of
  Table~\ref{table:clfs}.}
\label{fig:fitsz09}
\end{figure*}

Figures~\ref{fig:fitsz0} and~\ref{fig:fitsz09} show how well the
best-fit CLFs recover the observed XLF, $w(r_p)$ and $r_0$ in both
redshift bins. To illustrate the
reliability of the results, the $95$\%
confidence regions, which is a $\Delta \chi^2=12.59$ criterion for 6
degrees of freedom, is indicated in all
plots. The confidence regions are computed using a Monte-Carlo
approach. Beginning at the best-fit indicated in
Table~\ref{table:clfs}, one of the six parameters is increased or
decreased by at most $0.02$, and a new $\chi^2$ is calculated. If this
$\chi^2$ is within the confidence limit then the parameters are
stored, and a new random change to the parameters is made. After each
parameter has been randomly adjusted on average $500$ times, the
parameters are reset to the best-fit and the process is
restarted. This whole procedure is done 10 times for each fit,
resulting in $2569$ models within the confidence region at $z \approx
0$ and $2182$ models at $z \approx 0.9$. Then at $z \approx 0$, for each $L$ (in the
case of $\phi(L)$ or $r_0(L)$) or each $r$ (in the case of $w(r_p)$),
the CLF model is computed for each of the $2569$ set of parameters,
and the minimum and maximum $\phi(L)$, $r_0(L)$ and $w(r_p)$ are determined. These
minima and maxima determine the $95$\% confidence regions shown in the
figures. The same process is repeated to calculate the $z\approx 0.9$
confidence levels.

In addition to the XLF data, the $z\approx 0$ panel also shows the
best-fit XLF as determined by \citet{ball14} from fitting the XLFs in multiple
energy bands. This fit (as shown by the red long dashed line) can account
for all the local AGN demographics, including the \bat\ data. Although derived from a completely independent process, the
CLF-derived XLF (solid line) matches the \citet{ball14} measurement very well,
particularly around the knee of the XLF. This agreement gives us
confidence that the CLF model can be used to obtain accurate
information on the AGN population. 

\section{Results Derived from the AGN CLFs}
\label{sect:results}
In this section, we explore the statistics of various AGN physical quantities derived
from the CLF at these two redshifts, and discuss the implications for
our understanding of AGN evolution at different luminosities.

\subsection{AGN Bias and Mean Halo Mass}
\label{sub:bias}
\begin{figure*}
\centering
\includegraphics[angle=-90,width=0.45\textwidth]{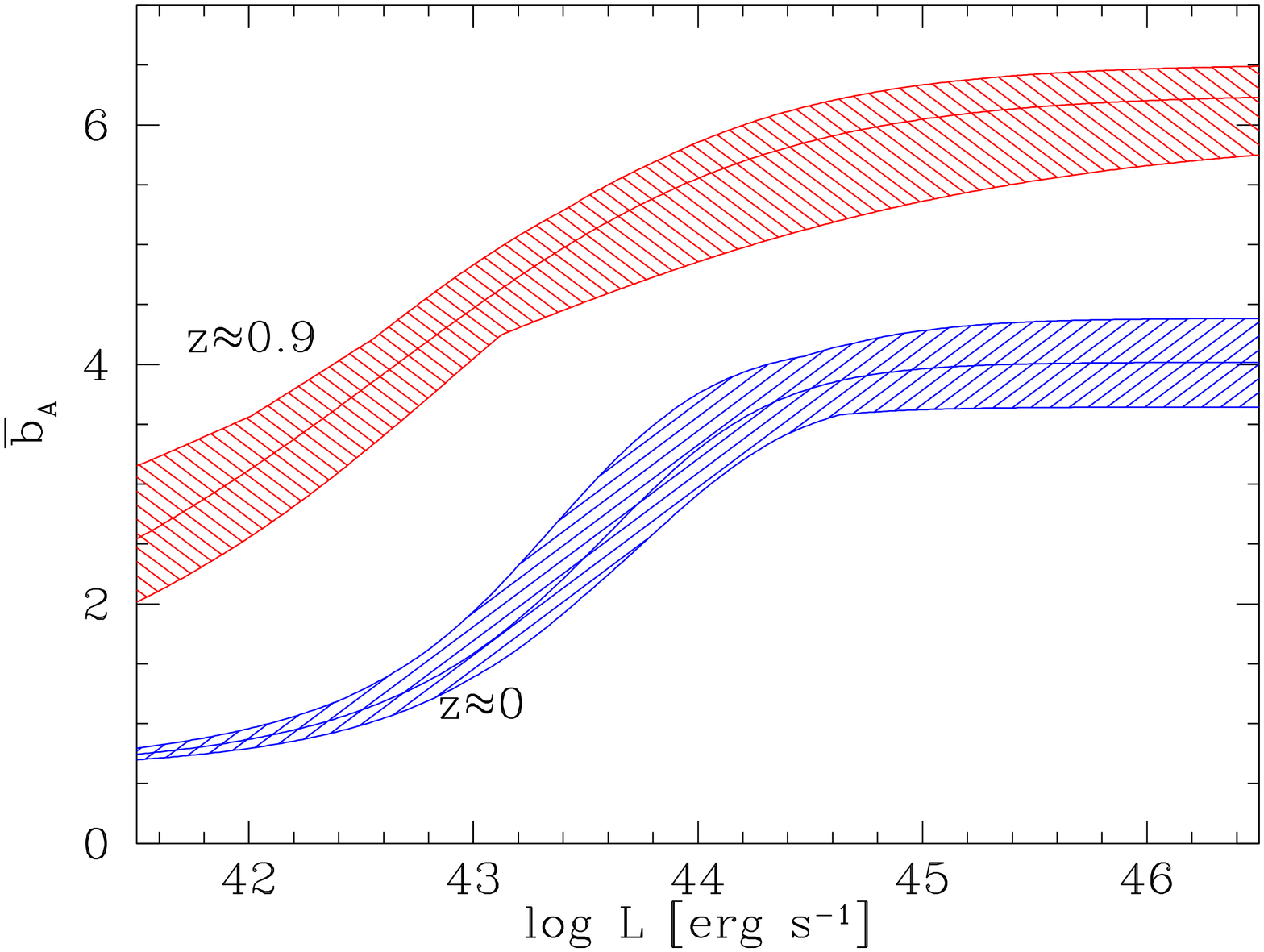}
\includegraphics[angle=-90,width=0.45\textwidth]{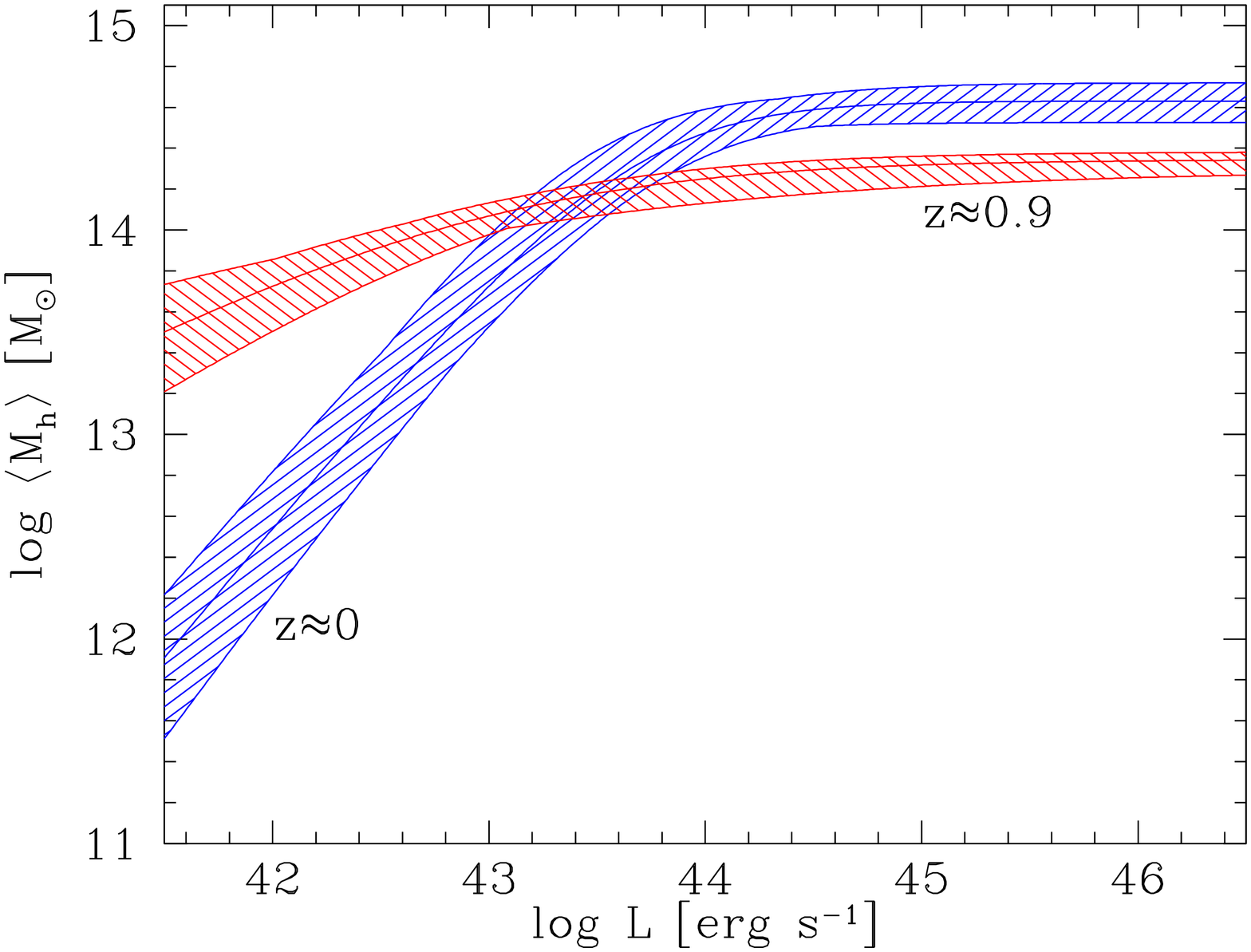}
\caption{The CLF-derived AGN bias, $\bar{b}_A$ (left), and mean halo mass,
$\langle M_{\mathrm{h}} \rangle$ (right), as a function of AGN X-ray
  luminosity at $z \approx 0$ (blue) and $z \approx 0.9$ (red). The solid lines are
  the predicted values, and the surrounding hatched areas are the 95\%
  confidence regions. Both quantities exhibit significant increases
  with luminosity, in particular at $\log (L/(\mathrm{erg\, s^{-1}})) \la 44$. This result
  likely indicates that different AGN fueling mechanisms are operating
at high and low luminosities. The luminosity dependence is weaker at
$z \approx 0.9$ which may be linked to a switch between merger-dominated
and secularly-triggered AGN growth \citep[e.g.,][]{db12}. The large
values of $\bar{b}_A$ and $\langle M_{\mathrm{h}} \rangle$ predicted
for local quasars result from the high-$L$ $r_0$ point of
\citet{cap10} and are therefore likely to be overestimated.}
\label{fig:biasmass}
\end{figure*}

The first two statistical properties of AGNs we will consider are the
mean AGN bias, $\bar{b}_A$ (Eq.~\ref{eq:bbar}), and the mean mass
of AGN-hosting haloes, $\langle M_{\mathrm{h}} \rangle$
(Eq.~\ref{eq:avgM}). Figure~\ref{fig:biasmass} plots the 95\%
confidence regions for these two quantities for both $z\approx 0$ and
$z\approx 0.9$.
There is a significant luminosity dependence to both quanitites at
each redshift that arises from the observed behaviour of $r_0(L)$. As
a result, for both $\bar{b}_A$ and $\langle M_{\mathrm{h}} \rangle$,
there seems to be a transition in behaviour at a luminosity of $\log (L/(\mathrm{erg\, s^{-1}}))
\approx 44$. Above that luminosity, the mass of AGN-hosting haloes is
approxminately constant, in agreement with surveys that show
little-to-no luminosity dependence in the clustering of quasars
\citep[e.g.,][]{shen13,eft15}. This indicates that these
high-luminosity AGNs have a
common triggering mechanism that depends only on being in a certain,
highly biased, environment. The wide span of luminosity over which the
mass is roughly constant can be
interpreted by invoking a range of black holes
masses and the fact that that all AGNs exhibit significant variability
\citep[e.g.,][]{hick14}. In contrast, at $\log (L/(\mathrm{erg\, s^{-1}})) \la 44$ AGN triggering
must be occurring by a process that depends on halo
mass. Moreover, the lower bias means that these AGNs do
not need to inhabit significantly dense environments and can therefore
be triggered by processes internal to the host halo, such as accretion from the halo itself
\citep[e.g.,][]{fan13}. Interestingly, the
$z \approx 0.9$ luminosity-dependence is weaker than at $z \approx 0$ which
likely means that there is a larger fraction of AGNs triggered in
massive haloes at
all luminosities during this epoch. Such envrionments are more
conducive to galaxy interactions and mergers, and gas-rich galaxies
are more common at higher redshifts, thus these results indicate that
AGN triggering by major mergers is more common at $z \sim 1$ and
becomes much rarer at low $z$ \citep[e.g.,][]{salucci99,db12}.

The average bias and mean halo mass of $z \approx 0$ quasar hosts
appears unrealistically large, as they imply local quasars should
reside in nearby massive clusters, which is not observed \citep[e.g.,][]{mart13}. The
large values result from fitting the high-$L$ $r_0$ point from
\citet{cap10} and therefore appears to be overestimated. That analysis
relied on the 3-year \bat\ catalog with only 199 AGNs. The latest
\bat\ catalog by \citet{baum13} contains over 600 AGNs, so an update to the
clustering analysis should give significant improvements to the local
CLF model. In contrast, the $z \approx 0.9$ results show that quasars
at that redshift might be commonly found in haloes with
$\log (M_{\mathrm{h}}/\mathrm{M_{\odot}}) \approx 14-14.3$,
corresponding to large groups or young clusters at this redshift. This
result will be discussed further in Sect.~\ref{sub:z09mass}.

\begin{figure}
\includegraphics[angle=-90,width=0.5\textwidth]{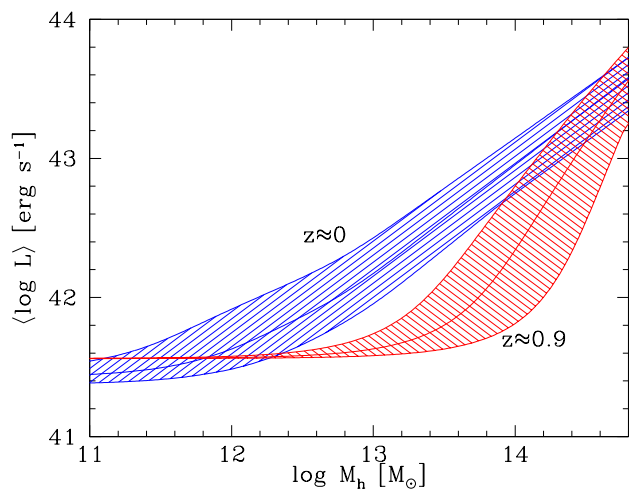}
\caption{The average AGN luminosity as a function of halo mass at $z
  \approx 0$ (blue line and hatched region) and $z \approx 0.9$ (red
  line and hatched region). In both
  cases, the average luminosity increases with mass, but only above a halo mass of $\log
  (M_{\mathrm{h}}/\mathrm{M_{\odot}}) \approx 12$. The slopes are also different
  ($\approx 1.2$ at $z\approx 0.9$; $\approx 0.8$ at $z \approx 0$),
  indicating that the local AGN population is more spread out in haloes of
  different masses.}
\label{fig:meanLvsM}
\end{figure}

A related view of these relationships can be obtained by calculating
the mean AGN luminosity as a function of halo mass, $\langle L
\rangle(M_{\mathrm{h}})$, via
\begin{equation}
\langle L \rangle(M_{\mathrm{h}}) = {\int L {d \Phi \over
  d\log L}(L,M_{\mathrm{h}}) d\log L \over \int {d \Phi \over
  d\log L}(L,M_{\mathrm{h}}) d\log L},
\label{eq:meanL}
\end{equation}
where
\begin{equation}
{d \Phi \over d\log L}(L,M_{\mathrm{h}}) = \Psi(L|M_{\mathrm{h}})
n(M_{\mathrm{h}}) dM_{\mathrm{h}}
\end{equation}
is the XLF for AGNs in haloes with mass
$M_{\mathrm{h}}$. Figure~\ref{fig:meanLvsM} plots the results for the
CLFs derived at $z \approx 0$ and $0.9$.
The mean AGN
luminosities at both $z \approx 0$ and $0.9$ increases with halo mass, as long as $\log (M_{\mathrm{h}}/\mathrm{M_{\odot}}) \ga
12$ (for $z \approx 0$) or $13$ ($z \approx 0.9$). The slopes of the
relations also differ substantially. At $\log (M_{\mathrm{h}}/\mathrm{M_{\odot}}) > 12.5$,
we find $L \propto M_{\mathrm{h}}^{0.8}$ at $z \approx 0$, although
this slope may be an overestimated due to the large \citet{cap10}
$r_0$ point causing high-$L$ AGNs to inhabit massive haloes. At $z
\approx 0.9$, the slope is measured for $\log
(M_{\mathrm{h}}/\mathrm{M_{\odot}}) > 13.5$, and we find $L
\propto M_{\mathrm{h}}^{1.2}$. The slope at $z
\approx 0$ appears to be in good agreement with the results of a similar analysis
by \citet{hutsi14}, but the $z \approx 0.9$ slope is much steeper than
their result at the same redshift. This difference most likely arises
from the assumption made by \citet{hutsi14} that the power-law
relation between luminosity and halo mass spans the entire range in
mass. Similarly, the slope of $1.2$ at $z \approx 0.9$ is not very
different from the slope of $1.4$ predicted by \citet{croton09}, but,
as the AGN triggering physics has changed,
the flatter slope at $z \approx 0$ is not consistent with the
\citet{croton09} model.

Figure~\ref{fig:meanLvsM} shows that at $z \approx 0.9$ the average
AGN luminosity is nearly constant with halo mass until $\log
 (M_{\mathrm{h}}/\mathrm{M_{\odot}}) \approx 13$. This means that in less massive haloes, the
average AGN luminosity is independent of the hosting halo mass,
consistent with either the fading of previously triggered quasars, or
a form of triggering by secular processes \citep{db12}. This behaviour
is roughly consistent with the instantaneous $z=1$ $L-M_{\mathrm{h}}$
relation found by \citet{chatt11} using a hydrodynamical simulation of
AGN evolution. Above $\log
M_{\mathrm{h}} \approx 13$, the mean AGN luminosity increases steeply
with halo mass, implying only these massive haloes can produce AGNs
that are, on average, typical of quasars. Interestingly, at $z \approx
0$ the average AGN luminosity depends on halo mass over a broad
range of masses. In the local Universe, AGNs can be found in a
variety of environments since they are likely triggered by a variety
of internal processes that may operate on a wide range of timescales. At both redshifts the mean AGN luminosity is
often significantly lower than the typical quasar value of $\log (L/(\mathrm{erg\, s^{-1}}))
\approx 44$; this, of course, follows from the XLF which shows that at
these redshifts, lower luminosity AGNs dominate the space density.

\subsection{AGN Halo Occupation}
\label{sub:occ}
\begin{figure}
\includegraphics[width=0.5\textwidth]{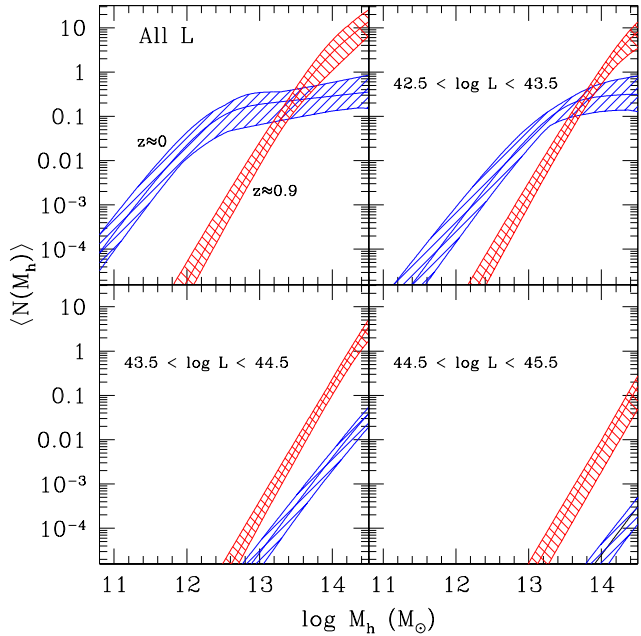}
\caption{The CLF-derived mean halo occupation number of AGNs at $z
  \approx 0$ and $0.9$ in four different luminosity ranges. The solid
  lines are the predicted values with the hatched regions denoting the
95\% confidence limits. As the CLF model only treats the `2-halo'
term of the correlation function, these predictions are unable to
distinguish between central and satellite AGNs.}
\label{fig:avgN}
\end{figure}

The average number of AGNs occupying haloes of different masses, $\langle N(M_{\mathrm{h}})\rangle$, allows
another view into the triggering and evolution of AGNs in their
cosmological environment. Figure~\ref{fig:avgN} shows the CLF
predictions for $\langle N(M_{\mathrm{h}})\rangle$ (computed from
Eq.~\ref{eq:avgN}) in four luminosity ranges at both $z \approx 0$ and
$0.9$.
Recall that the AGN CLF was constrained using only that part of the correlation
function dominated by the `2-halo' term, so the predicted $\langle
N(M_{\mathrm{h}})\rangle$ contains contributions from both central and
satellite AGNs. The satellites will only become relevent for haloes
above $\log (M_{\mathrm{h}}/\mathrm{M_{\odot}}) \sim 13.5$ (i.e., group scale or larger;
\citealt{chatt11,lea15}), and, indeed, it is only for those masses $\langle
N(M_{\mathrm{h}})\rangle > 1$.

At $z \approx 0$, the total AGN occupation number rises quickly from
$\log (M_{\mathrm{h}}/\mathrm{M_{\odot}}) \approx 11$ to $12.5$ before flattening to a much shallower
rise to larger halo masses. In the local Universe, once they are
integrated over luminosity, AGNs are found in
haloes with $\log (M_{\mathrm{h}}/\mathrm{M_{\odot}}) \ga 12.5$ at
roughly equal numbers (roughly, $0.2$--$0.5$ AGNs per halo). This shape is
roughly consistent with the results of \citet{lea15} who estimate the
AGN halo occupation at $0.2 < z < 1$ with $41.5 \la \log (L/(\mathrm{erg\, s^{-1}})) \la
43.5$. However, considering the AGN population in finer $L$ bins yields
occupation distributions that, by construction, are best described as
cutoff power-laws. Following the XLF at $z \approx 0$, the occupation
number falls quickly with luminosity with quasars occupying only a
tiny fraction of haloes.

The shape of the $z \approx 0.9$ AGN occupation curves are similar to 
the ones at $z \approx 0$, but, following the evolution in the XLF, the
curves are shifted so that more luminous AGNs are more
numerous. The occupation for all AGNs at $z \approx 0.9$
is greater than unity for haloes at $\log (M_{\mathrm{h}}/\mathrm{M_{\odot}}) \ga
13.8$. These AGNs are mostly luminous Seyferts and quasars, and are
found in haloes which are in the process of collapsing
\citep{van02}. Clearly, processes in these haloes allow efficient
triggering of AGNs that is not found in their lower redshift analogues
(Sect.~\ref{sub:z09mass}). 

Earlier measurements of the AGN halo occupation find that the shape
of the distribution for `central' AGNs is a step or an error function
\citep{chatt11,miy11,lea15}, and that the satellite population is best
described by a cutoff power-law. In contrast, we find that the
occupations distribution of all AGNs is consistent with a cutoff power-law
\emph{when condidered over a narrow luminosity range}. After
integrating over luminosity and summing up the individual cutoff
power-laws, we find a AGN halo occupation distribution that is flatter
and more consistent with the step and error functions found by
previous authors. This result illustrates the importance of
considering the luminosity dependence of the AGNs when describing the
halo occupation statistics. 

\subsection{Average Black Hole Growth Rate}
\label{sub:growth}
The upper-left panel of Fig.~\ref{fig:avgN} shows the mean number of
AGNs of all luminosity in dark matter haloes as a function of halo
mass. Multiplying this curve by the average luminosity of AGNs in each
AGN-occupying halo (Fig.~\ref{fig:meanLvsM}) gives the AGN
luminosity averaged over all haloes of a given mass. This quantity
therefore gives the average accretion-driven black hole growth rate as
a function of halo mass at the given $z$.

\begin{figure}
\includegraphics[angle=-90,width=0.5\textwidth]{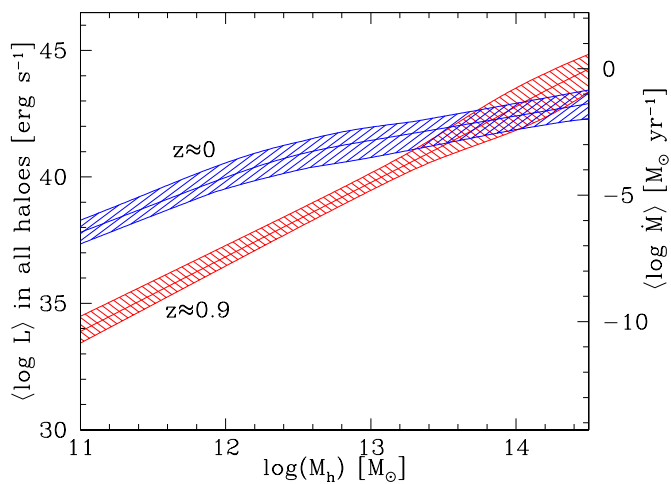}
\caption{Average AGN luminosity produced by all dark matter haloes as
  a function of halo mass. The right-hand axis translates this to an
  average black hole growth rate in haloes after assuming a radiative
  efficeincy of $0.1$ and a bolometric correction of $30$
  \citep{vf09}. At $z\approx 0.9$ the vast majority of the black hole
  growth occurs in high mass haloes. However, the black hole growth in
lower mass haloes increases by several orders of magntidue from
$z\approx 0.9$ to $\approx 0$, consistent with the concept of cosmic
downsizing.}
\label{fig:growth}
\end{figure}

Figure~\ref{fig:growth} shows the results of this calculation for both
redshifts where
the average AGN X-ray luminosity is indicated on the left-hand axis
and the equivelant average black hole accretion rate (assuming a
radiative efficiency of $0.1$ and a bolometric correction of $30$;
\citealt{vf09}) is noted on the right-hand axis.
There is a marked
difference between the two epochs. At $z\approx 0.9$, black hole
growth is strongly weighted to higher mass haloes, where both the
occupation number and mean luminosity of AGNs are large, while black
holes in haloes with masses $\log (M_{\mathrm{h}}/\mathrm{M_{\odot}}) \la
13$ are growing very slowly. It seems that the turbulent
environment of a collapsing $\log (M_{\mathrm{h}}/\mathrm{M_{\odot}}) \ga
14$ halo provides numerous mechanisms for the galaxy interactions
necessary for the growth of the massive black holes seen in galaxies
within nearby clusters. At $z \approx 0$, the growth rate in the
lower-mass haloes has risen by several orders of magntidues, and the
rate in the high mass haloes has fallen. Overall, this figure nicely
illustrates the 'cosmic downsizing' \citep[e.g.,][]{has05,bundy06,ueda14} concept where the most massive
black holes and galaxies are grown earlier on in the Universe before
the lower mass black holes and galaxies. 

\subsection{AGN Lifetime}
\label{sub:lifetime}
As mentioned in Sect.~\ref{sub:CLFintro}, the AGN CLF can be used to
estimate the AGN lifetime as a function of luminosity by computing the
fraction of haloes that host AGNs at a given luminosity. Since there is
a one-to-one correspondance between luminosity and mean halo mass
(Fig.~\ref{fig:biasmass}), each luminosity corresponds to a mean mass
from which the space density of haloes at that mass can be computed
(Eq.~\ref{eq:time}). This estimate of lifetime assumes the halo
lifetimes are equal to the Hubble time, which means the AGN lifetimes
are, strictly speaking, upper-limits.   

\begin{figure}
\includegraphics[angle=-90,width=0.5\textwidth]{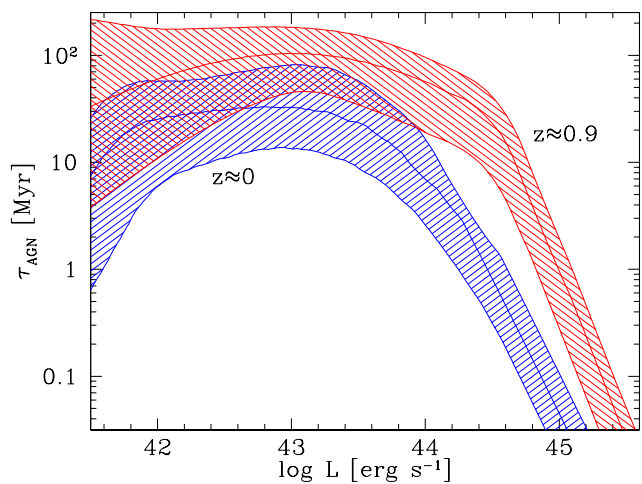}
\caption{AGN lifetimes at $z \approx 0$ and $0.9$ estimated from
  Eq.~\ref{eq:time} and the best-fitting CLF models. The hatched
  regions denote the 95\% confidence levels around the predicted
  lifetimes (solid lines). As with the other statistics, there is a
  strong luminosity dependence to the lifetimes, with lower luminosity
AGNs exhibiting longer lifetimes than the higher luminosity quasars.}
\label{fig:time}
\end{figure}

The lifetimes computed from our CLF models are shown in
Figure~\ref{fig:time}, with the hatched regions showing the 95\%
confidence regions.
The lifetimes generally fall in the range of $10^7$--$10^8$~yrs,
except at the highest luminosities. Depending on how one
defines quasar, our quasar lifetimes are in good agreement with the
predictions from the semi-analytical models of \citet{croton09} and
\citet{cw13} as well as the range of estimates found from a variety of
methods \citep{mart04}. At both redshifts, the lifetimes exhibit a significant luminosity
dependence, with the lifetimes falling at high luminosities. This
shape follows from the shape of the XLF and is consistent with other
estimates of the quasar lifetime \citep{croton09}. At
lower-luminosities, more consistent with Seyfert galaxies, the
lifetimes are roughly constant with luminosity. The transition in lifetimes again indicates the
two modes of AGN fueling physics that operates at low and high
luminosity.

\subsection{How to Populate Haloes with AGNs in Numerical Simulations}
\label{sub:simulations}
Despite rapid progress in technology, cosmological simulations
continue to have difficulty accurately capturing the detailed physics
of AGN fueling and feedback because of the small size and time scales
that must be resolved during the calculation
\citep[e.g.,][]{spring05,si15}. As a result, there remains
considerable uncertainty in how well the simulated AGN population
resembles the actual one, particularly with regards to AGN fueling,
appearance (e.g., obscured versus unobscured) and lifetimes
\citep[e.g.,][]{hkb14,cs15}. This uncertainty ultimately impacts
theories of how AGN activity affects galaxy evolution. Many
cosmological calculations use the technique of 'abundance matching'
\citep[e.g.,][]{cwk06,croton09} to ensure the AGN luminosity function is followed; however, how
these AGNs are assigned to different halo masses or given luminosities
is extremely model dependent. The CLF methodology developed here, which naturally gives
relationships between luminosity and halo mass (e.g., Fig.~\ref{fig:biasmass}), provides a more observationally
motivated way of assigning AGNs to numerical haloes. Below, we use the
$z\approx 0.9$ AGN CLF derived above to illustrate
three ways the CLF can assist cosmological simulations in
their treatment of AGNs.

\begin{figure}
\centering
\includegraphics[width=0.45\textwidth]{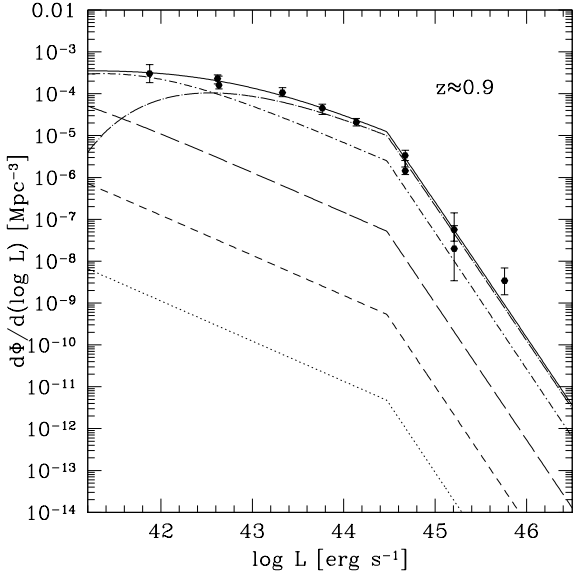}
\caption{The contribution to the $z
  \approx 0.9$ XLF from AGNs hosted by dark matter
  haloes of various masses. The data points and the solid line are the same as in
  Fig.~\ref{fig:fitsz09}. The other lines
  indicate the XLF from AGNs hosted by haloes with $\log (M_{\mathrm{h}}/\mathrm{M_{\odot}})
< 11$ (dotted), $11 < \log (M_{\mathrm{h}}/\mathrm{M_{\odot}}) < 12$ (short dashed), $12 <
\log (M_{\mathrm{h}}/\mathrm{M_{\odot}}) < 13$ (long dashed), $13 < \log (M_{\mathrm{h}}/\mathrm{M_{\odot}}) <
14$ (short dash-dotted), and $14 < \log (M_{\mathrm{h}}/\mathrm{M_{\odot}})$ (long
dash-dotted).}
\label{fig:xlfmhs}
\end{figure}

The first possibility is
to update the idea of abundance matching by considering the halo mass
dependence of the XLF. This is shown in Figure~\ref{fig:xlfmhs} where
the derived CLF at $z \approx 0.9$ is used to compute XLFs for
different ranges of \mh. 
This plot encapsulate many of the results presented above; namely,
the transition in halo host mass at $\log (L/(\mathrm{erg\, s^{-1}})) \sim 44$. These
halo-dependent XLFs will provide crucial new information to modelers
when populating AGNs in numerical haloes by matching the observed
luminosity function.

\begin{figure}
\centering
\includegraphics[width=0.5\textwidth]{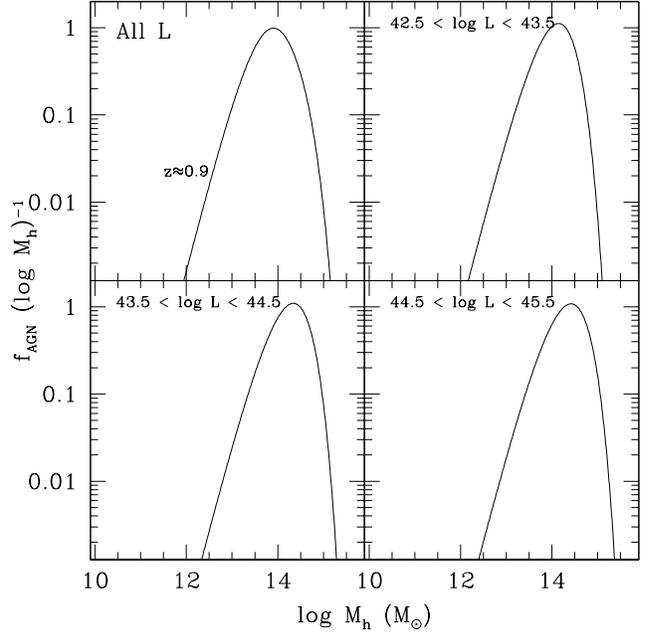}
\caption{AGN halo mass probability density function (Eq.~\ref{eq:halopdf}) in
  different luminosity bins at $z
\approx 0.9$. This type of plot can be used to populate AGNs with the
appropriate luminosities in dark matter haloes in cosmological simulations.}
\label{fig:halopdf}
\end{figure}

An alternative way of doing the assignments of AGNs numerically is through probabilities, and by
using the CLF we can calculate the AGN halo mass probability density function \citep{lea15}, defined
as
\begin{equation}
\label{eq:halopdf}
f_{\mathrm{AGN}} = {\left < N(M_{\mathrm{h}}) \right > n(M_{\mathrm{h}})
  \over \int \left < N(M_{\mathrm{h}}) \right > n(M_{\mathrm{h}})
dM_{\mathrm{H}} }.
\end{equation}
This quantity, which is plotted for $z \approx 0$ and $0.9$ in
Figure~\ref{fig:halopdf}, shows how probable it is for AGNs to be
hosted by haloes of a given mass, and can be used in conjunction with
the traditional abundance matching technique.
As long as the computed halo mass function is compatible with the
\citet{tink10} used here, then populating haloes with these probability
density functions should automatically reproduce the AGN XLF at the
appropriate redshift.

\begin{figure}
\includegraphics[angle=-90,width=0.45\textwidth]{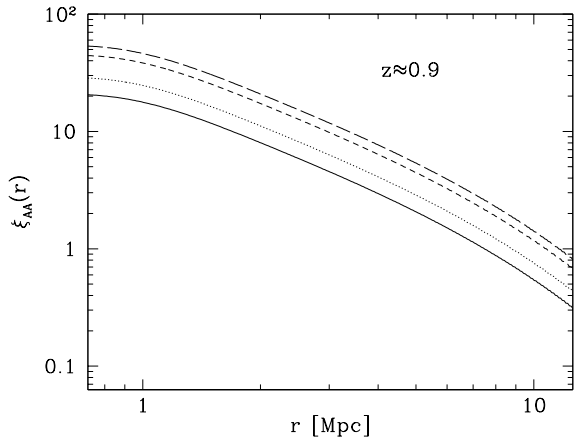}
\caption{Predictions for the AGN 2-point correlation function,
  $\xi_{\mathrm{AA}}(r)$, at $z \approx 0.9$. The different curves differentiate between AGNs at
  different X-ray luminosities: $42.5 < \log (L/(\mathrm{erg\, s^{-1}})) < 43.5$ (dotted), $43.5
  < \log (L/(\mathrm{erg\, s^{-1}})) < 44.5$ (short-dashed), $44.5 < \log (L/(\mathrm{erg\, s^{-1}})) < 45.5$ (long-dashed),
  and integrated over all $L$ (solid). The XLF of AGNs shows that
  the space density of AGNs is dominated by lower-luminosity objects,
  so the integrated $\xi_{\mathrm{AA}}(r)$ is closer to the curve for
the lower-luminosity bin. These are `2-halo' correlation functions,
and are therefore cutoff at small seperations.}
\label{fig:AGNcf}
\end{figure}

The luminosity-dependent AGN correlation function is a further
constraint to the AGN population that has not often been used when
populating haloes. In Fig.~\ref{fig:AGNcf}, the CLF-predicted
$\xi_{\mathrm{AA}}(r)$ for $z \approx 0.9$ AGNs in different luminosity bins are
plotted.
As found by observations \citep{kou13}, the correlation length
increases with luminosity, and this behavior is captured by the
CLF. Combining these correlation functions with either of the two
changes to the abundance matching techniques described above will allow cosmological
models to more accurately include the effects of AGNs in the growth
of galaxies in a more observationally-rigorous manner. For example,
once the CLF results are used to populate AGNs in haloes at a specific
$z$, the simulation can be used to examine the history or future of
those haloes, and make predictions for how the clustering and fueling
of those AGNs have changed or will change with time.

\section{Discussion}
\label{sect:discuss}

\subsection{AGNs in their Cosmological Environments:
  Luminosity-Dependence}
\label{sub:lumindep}
The central aspect of the CLF methodology is that it naturally
produces a luminosity-dependent AGN HOD model. However, measurements
of $r_0(L)$ are necessary in order to constrain the CLF, and therefore
the predictions derived from the CLF are limited by the number and
accuracy of these measurements. Despite this limitation, there is
substantial interest in whether the environment of AGN host galaxies
is connected to AGN observational characteristics such as luminosity
and obscuration. The CLF is a valuable tool to examine these
questions, especially as further $r_0(L)$ measurements become
available over the next decade.

We have decided to use X-ray survey data to constrain and study the
AGN CLF despite the relatively limited number of datasets available
compared to studies of optical quasars. This was done because X-ray
surveys provide a view of the AGN population that is both unbiased to
Compton-thin obscuration and covers a substantial range in
luminosity. Optical quasar surveys are effective in sampling the
high-$z$, unobscured AGN population, but are typically sensitive to a
narrower range in luminosity than what is found in X-ray surveys
\citep[e.g.,][]{krumpe15}. As we are specifically interested in examining the luminosity
dependence of the host halo properties, the X-ray data provides the
best constraints. However, as pointed out by \citet{lea15}, the data
products produced by X-ray AGN surveys may also be biased to more
massive host galaxies, as only these will be luminous enough to obtain
a redshift measurement. This effect may influence the inferred masses
of AGN hosting haloes.

Bearing this caveat in mind, the results of the previous section shows
that the AGN luminosity does depend on the halo properties at both $z
\approx 0$ and $0.9$. This luminosity dependence grows stronger as the
Universe ages, and can be reasonably interpreted as demonstrating two
different modes of AGN triggering. One mode can generate short-lived
high-luminosity quasars at any $z$; these objects are rare and therefore
typically populate high mass haloes, but, if the environment is
correct, the AGN luminosity does not depend on the halo mass. These
properties roughly translate into a merger triggered AGN scenario
\citep[e.g.,][]{lidz06,fan13}. The other mode of AGN triggering becomes prominant at
$z \la 0.9$ and produces lower-luminosity AGNs that are much
longer-lived, but with luminosities that depend on halo mass. These
lower-luminosity AGNs can also be found over a range of different halo
mass, and may be triggered by 'secular' processes (including disk
instabilities, minor mergers, galaxy harassment, halo-gas accretion)
that can operate in a wide range of cosmological environments. 

The luminosity-dependence of our results makes it challenging to
perform quantitative comparisons to previous work. Moreover,
differences in assumed cosmology and the $\bar{b}_{\mathrm{A}}
\rightarrow M_{\mathrm{h}}$ conversion lead to additional errors when
performing comparisons. Therefore, we will focus on a general
comparison between our results and those from the literature. Common
values for halo masses found from clustering of X-ray selected AGNs
are $\log (M_{\mathrm{h}}/\mathrm{M_{\odot}}) \sim 13-13.5$ (independent of $z$;
\citealt{caf12}) which is roughly consistent with the CLF-derived
results for lower-luminosity AGNs at $z \approx 0.9$, but misses the strong luminosity
dependence at $z \approx 0$. \citet{lea15} estimate halo masses of moderate luminosty
low-$z$ X-ray AGNs using lensing information and find halo masses
$\sim 10^{12.5}$~M$_{\odot}$, which is in rough agreement with our
results. Recently, \citet{mendez16} measured the clustering of X-ray
AGN in two luminosity bins, one with $\langle \ \log (L/(\mathrm{erg\,
  s^{-1}})) \rangle \sim 42.4$ and $\langle z \rangle =0.57$, and the
other with $\langle \ \log (L/(\mathrm{erg\,
  s^{-1}})) \rangle \sim 43.2$ and $\langle z \rangle =0.87$. For the
latter sample, they find $\log (M_{\mathrm{h}}/\mathrm{M_{\odot}})
\sim 13.4$ if they include the COSMOS field, and $13.1$ if they omit
COSMOS, values that are $\approx 4\times$ lower than the CLF predictions
(Fig.~\ref{fig:biasmass}). In contrast, the lower $z$ and luminosity
sample of \citet{mendez16} has
about the same halo mass and bias as the higher luminosity bin, a
result that disagrees with the expectation from the CLF (and the
\citet{lea15} results).

Interestingly, optically-selected quasars seem to populate haloes with
masses about an order of magnitude lower than X-ray select AGNs
\citep{caf12}. However, as argued by both \citet{lea15} and \citet{mendez16}
comparing clustering results between AGNs selected at different
wavelength leads to strong biases, since the host galaxies properties
can be very different between the two samples. For example, as mentioned
above, X-ray selected samples may be biased to higher mass galaxies
(and hence higher mass haloes) due to the need to obtain host galaxy
redshifts. Optical quasars are likewise subject to biases related to
the detection of strong emission lines with the correct ratios
\citep[e.g][]{moran02,trump15}. Thus, a straightforward comparison
between the two sets of results is not possible. Future CLF
modeling will focus on creating a connection from the X-ray based
method to the existing optical clustering measurements at higher
redshifts that will test the effects of these biases.

\subsection{Luminous AGNs in High Redshift Young Galaxy Clusters}
\label{sub:z09mass}
One of the most striking predictions of the AGN CLFs derived here is
the cluster-sized masses for haloes hosting luminous AGNs ($\log
(L/(\mathrm{erg\, s^{-1}})) \ga 43$). This prediction is not supported
by observations at $z \approx 0$, where luminous AGNs are very rarely
found in galaxy clusters, and, indeed, the cluster AGN fraction is
smaller than the field fraction \citep{gal09,ele13,mart13}. Thus, even though the halo
mass found here agrees with the one inferred by \citet{cap10}, the
large value of $r_0$ used to constrain the CLF at high luminosity is
likely overestimated and therefore giving erroneous results at these
luminosities. An update to the \bat\ clustering analysis would
significantly clarify the results for local AGNs.

The situation at $z \approx 0.9$ is significantly different. In
fact, there are several lines of evidence that luminous AGNs in $\log
(M_{\mathrm{h}}/\mathrm{M_{\odot}}) \sim 14$ haloes are common at $z
\sim 1$. First, the fraction of luminous AGNs in clusters rises
rapidly with redshift to the point where the fraction appears to
be at least
equal to that of the field at $z \ga 1$ \citep{gal09,mart13,albert16,buf16}. In addition, there are
examples where deep \chandra\ observations of $z \sim 1$ quasars are
finding hot cluster gas emission \citep{sie10,russell12,julie16}. In fact, \citet{russell12} suggests the
possibility that high-$z$ cool-core X-ray clusters may be associated
with many quasars, but are not-identified due to the glare of the
quasar. Lastly, in a study of the AGN emission in BCGs, \citet{hl13} finds
that these galaxies become significantly more radiatively efficient
and luminous at higher redshifts and could host quasars at $z \sim
1$. Taken together, these results all suggest that luminous AGN
activity at $z \approx 0.9$ may indeed be common, if not dominant, in $\log
(M_{\mathrm{h}}/\mathrm{M_{\odot}}) \ga 14$ haloes. Such a result
would be consistent with the major-merger triggering mechanism for AGNs, as
these young cluster environments at high redshifts would allow more frequent
interactions between gas-rich galaxies than at lower redshift. As
the clusters continue to grow and relax, their velocity disperions
increase and interactions become rarer, rapidly reducing the number of
luminous AGNs in clusters to the low values observed today \citep[e.g.,][]{ele15}.

If this scenario is accurate, the question arises of why do the
previous X-ray clustering analyses all tend to give halo masses of
$\log (M_{\mathrm{h}}/\mathrm{M_{\odot}}) \sim 13-13.5$ at almost all
$z$. The main difference between the CLF method and more traditional
cluster techniques is that the CLF considers the luminosity variation
in the clustering, and makes use of the total number density of AGNs
at different luminosities and dark matter haloes at different
masses as additional constraints. This additional information makes
sure that the AGNs are distributed over the range of dark matter halo
masses in a way consistent with how $r_0$ varies with $L$. 

The CLF predictions for the dark matter hosts of quasars is partially based on
extrapolations from clustering measurements of lower-luminosity AGNs
\citep[e.g.,][]{kou13}. Therefore, the additional clustering data of $z
\sim 1$ quasars that will be obtained by \textit{eROSITA} will be
required to refine the CLF model and explore these predictions in
detail. 

\section{Summary and Conclusions}
\label{sect:concl}
This paper developed a method to constrain the AGN Conditional Luminosity
Function (CLF), a luminosity-dependent
halo occupation model for AGNs, thereby allowing a connection between AGN
properties and its cosmological environment. The method requires the
X-ray luminosity function and $r_0(L)$ over a narrow range of
redshifts. Using the limited data currently available, a CLF model was constrained at two redshifts, $z \approx 0$ and
$0.9$, and the AGN bias, mean halo mass, AGN lifetime, and halo
occupation numbers were all calculated as functions of luminosity. Our
most important results are:

\begin{itemize}
\item The AGN bias and mean halo mass show significant luminosity
  dependence at both redshifts, specifically for $\log
  (L/(\mathrm{erg\, s^{-1}})) \la 44$. This likely indicates a change
  in AGN fueling processes from low to high luminosity.  

\item In contrast to earlier clustering analyses, the CLF method
  predicts that high luminosity AGNs at $z \approx 0.9$ inhabit haloes with
  $M_{\mathrm{h}} \sim 10^{14}$~M$_{\odot}$, which are the
  ancestors of local massive clusters. This result is consistent with
  the observed rapid increase with redshift of the luminous AGN
  fraction within clusters. These environments provide the right
  setting for strong interations between the massive gas-rich galaxies
  needed to fuel luminous nuclear accretion. Additional data from
  \textit{eROSITA} is needed to further explore this prediction.

\item The mean AGN luminosity depends on halo mass at both epochs, but
  only for $M_{\mathrm{h}} \ga 10^{12}$~M$_{\odot}$ ($z \approx 0$) or
  $M_{\mathrm{h}} \ga 10^{13}$~M$_{\odot}$ ($z \approx 0.9$). We find
  $L \propto M_{\mathrm{h}}^{0.8}$ at $z \approx 0$ and $L \propto
  M_{\mathrm{h}}^{1.2}$ at $z \approx
  0.9$ indicating that AGNs are spread over a wider
  mass range for each luminosity at low redshift. 

\item AGN lifetimes are $\la 10^8$~yrs for all luminosities and
  redshifts, and are $< 10^6$~yrs for the highest luminosities.

\item The CLF-derived results strongly support the idea that at $z \la
  1$ AGN triggering evolves from a merger-dominated regime which is
  only weakly dependent on halo mass to one where the majority of AGNs
  are triggered by secular processes, including interactions with
  neighbours \citep[e.g.,][]{koss10,db12,hkb14}. The high luminosity quasars must still be fueled
  by a catastrophic interaction with another gas rich galaxy \citep[e.g.,][]{san88},
  but such events are rare in the nearby Universe.

\item As the CLF method provides a statistical connection between AGN
  luminosity and host halo masses, it can be used to populate dark
  matter haloes in cosmological simulations with AGNs of
  different luminosities in a logical and
  self-consistent manner.

\end{itemize}

The CLF method described here relies on the accuracy of the underlying
observations. If certain datapoints are biased in some manner (such as
the high-$L$ $r_0$ point at $z \approx 0$; Fig.~\ref{fig:fitsz0}) then
the CLF-derived statistics will inherit that bias. However, as observations
improve with future observatories and campaigns, the CLF method can
quickly incorporate the new data and produce increasingly more
accurate predictions. For example, recent work by \citet{mount16}
indicate that $r_0(L)$ may decrease at $\log L \sim 43.5$ (see also
the models of \citealt{fan13}). If this result is confirmed by future
observations, the CLF framework can account for it by simply revising the underlying
CLF parameterization (Eq.~\ref{eq:clfform}). In general, future
$r_0(L)$ measurements in the X-ray band will provide the most
important improvement to the CLF-derived results. 

Future development of the CLF method will occur in two directions. As
mentioned earlier, the
first direction will focus on including the results from the optical quasar
clustering measurements. Additionally, we will also develop a model for the
  1-halo term in the AGN correlation function. The 1-halo term dominates at separations $\la 1$~h$^{-1}$Mpc and
arises from pairs of AGNs within a single halo (e.g., a central and
satellite galaxy). Therefore, it encodes information about AGN
triggering on small scales (e.g., mergers and other forms of galaxy
harassment). The addition of this term in the method will also allow a
connection to more detailed AGN host galaxy properties (e.g., stellar
mass) that also depend on position within haloes. 

\section*{Acknowledgements}
The author thanks J.\ Wise, G.\ Altay, A.\ Myers, and R.\ Grissom for
help and advice during the course of this work, R.\ Hickox for
comments on a draft of the manuscript, and acknowledges
support from NSF award AST 1333360.








\appendix

\section{Details on the Cosmological Calculations}
\label{app:cosmo}
First, define a dark matter halo of mass \mh\ as
\begin{equation}
\label{eq:halo}
M_{\mathrm{h}} = {4 \over 3} \pi R_{\mathrm{h}}^3 \bar{\rho}_m(z)
\Delta,
\end{equation}
where $R_{\mathrm{h}}$ is the radius of the halo, and $\Delta=200$
quantifies the overdensity of the halo compared to the average density
of the Universe; i.e., $\Delta \bar{\rho}_m= \Delta
\rho_{\mathrm{crit}} \Omega_m(z)$ ($\rho_{\mathrm{crit}}$ is the
critical density). 

With \mh\ now specified, the fitting formulas computed by \citet{tink10} are used to
compute \hmf\ and $b(M_{\mathrm{h}})$ at the required redshifts. In order to use
these formulas, the smoothed variance of the linear density field is
needed:
\begin{equation}
\sigma^2(r)={1 \over 2\pi^2} \int P(k)
\hat{W}^2(kr),
\end{equation}
where
\begin{equation}
P(k,z)=P_i(k)T^2(k)D^2(z)
\end{equation}
is the linear power spectrum and
\begin{equation}
\hat{W}(kr)= {3 \over  (kr)^3}(\sin(kr) -kr\cos(kr))
\end{equation}
is the Fourier transform of the top-hat window function in real
space. The shape of the power spectrum of fluctuations is initially
assumed to be $P_i \propto k$ and then is modified using the
transfer function $T(k)$ fitting formula provided by \citet{eh98} and the
growth function $D(z)$ \citep{cpt92}. The normalization of $P(k)$ is set by
using the local value of $\sigma$ for spheres of radius 8~$h^{-1}$Mpc,
$\sigma_8$.

The calculation of $\xi_{\mathrm{dm}}^{2h}$ requires computing both
the `1-halo' and total dark matter correlation functions. The latter
is computed from the non-linear evolved power spectrum
$P_{\mathrm{NL}}(k)$ via a Fourier Transform,
\begin{equation}
\xi_{\mathrm{dm}}(r)= {1 \over 2 \pi^2} \int_0^{\infty} k^3
P_{\mathrm{NL}}(k) {\sin kr \over kr} {dk
  \over k}.
\end{equation}
The \textsc{nicea} code \citep{kilb09} is used to calculate $P_{\mathrm{NL}}$
at the needed $z$ using the \citet{smith03} fitting function and the \citet{eh98}
transfer function. The resulting $\xi_{\mathrm{dm}}(r)$ closely follow
the results from the Millennium II simulation \citep{mbk09} at radii $r \ga
1$~$h^{-1}$Mpc (after correcting for the difference in $\sigma_8$). 

The `1-halo' term of the correlation function is \citep{vym03}
\begin{equation}
\xi_{\mathrm{dm}}^{1h}(r)=\int_0^{\infty} dk k^2 {\sin kr \over kr}
\int_0^{\infty} dM_{\mathrm{h}} n(M_{\mathrm{h}})
  |\hat{\delta}(M_{\mathrm{h}},k)|^2,
\end{equation}
where $\hat{\delta}(M_{\mathrm{h}},k)$ is the Fourier Transform of the
  halo density profile truncated at the virial radius. Here, a
  Navarro-Frenk-White (NFW; \citealt{nfw97}) density profile is assumed for the
  halo and the formulas for computing $\hat{\delta}(M_{\mathrm{h}},k)$
    are given by \citet{scocc01}. The evolution of the halo concentration
    parameter with $z$ and \mh\ is calculated using the relation of
    \citet{zcz07}. The `2-halo' term is then computed from
    $\xi_{\mathrm{dm}}^{2h}(r)=\xi_{\mathrm{dm}}(r)-\xi_{\mathrm{dm}}^{1h}(r)$
    assuming that $\xi_{\mathrm{dm}}(r)=\xi_{\mathrm{dm}}^{1h}(r)$ at
    $r=0.01$~Mpc.

\section{Light-Cone Effects on the CLF}
\label{app:light-cone}
In a situation where the data constraining the AGN CLF is accumulated
over a significant redshift range than a more accurate CLF might be
determined by averaging \hmf\ and $\xi_{\mathrm{dm}}^{2h}(r)$ over
$z$. For example, if the AGN data is gathered between
$z_{\mathrm{min}}$ and $z_{\mathrm{max}}$, then the XLF would be
\begin{equation}
\phi(L)=\int_{0}^{\infty} \Psi(L|M_{\mathrm{h}}) n_{\mathrm{eff}}(M_{\mathrm{h}})
dM_{\mathrm{h}},
\end{equation}
where
\begin{equation}
n_{\mathrm{eff}}(M_{\mathrm{h}}) = {1 \over V}
\int_{z_{\mathrm{min}}}^{z_{\mathrm{max}}} {dV \over dz}
n(M_{\mathrm{eff}},z) dz
\end{equation}
and
\begin{equation}
V=\int_{z_{\mathrm{min}}}^{z_{\mathrm{max}}} {dV \over dz} dz.
\end{equation}
Similarly, the AGN correlation function is now
\begin{equation}
\xi_{\mathrm{AA}}^{2h}(r) \approx \bar{b}_{\mathrm{A}}^2
\xi_{\mathrm{dm,eff}}^{2h}(r),
\end{equation}
where
\begin{equation}
\xi_{\mathrm{dm,eff}}^{2h}(r)={\int_{z_{\mathrm{min}}}^{z_{\mathrm{max}}}
  {dV \over dz} n^2(M_\mathrm{h},z) \xi_{\mathrm{dm}}^{2h} dz \over \int_{z_{\mathrm{min}}}^{z_{\mathrm{max}}}
  {dV \over dz} n^2(M_\mathrm{h},z) dz}
\end{equation}
and
\begin{equation}
\bar{b}_A={1 \over \phi(L)} \int_0^{\infty} \Psi(L|M_{\mathrm{h}})
b_{\mathrm{h}}(M_{\mathrm{h}}) n_{\mathrm{eff}}(M_{\mathrm{h}})
dM_{\mathrm{h}}.
\end{equation}
In these equations, $dV/dz$ is the comoving volume element per unit
solid angle.



\bsp	
\label{lastpage}
\end{document}